\documentclass[copyright,creativecommons]{eptcs}
\providecommand{\event}{SYNT 2016}

\usepackage[top=3cm,bottom=3cm,left=3cm,right=3cm]{geometry}
\usepackage[characters]{ltl}
\usepackage{listings}
\usepackage{verbatim}
\usepackage{url}
\usepackage{multirow}
\usepackage{xspace}
\usepackage{float}

\colorlet{codecolor}{blue!70!black}

\newcommand{\bools}{\ensuremath{\mathbb{B}}}
\newcommand{\nats}{\ensuremath{\mathbb{N}}}
\newcommand{\enums}{\ensuremath{\mathbb{E}}}
\newcommand{\buses}{\ensuremath{\mathbb{U}}}
\newcommand{\temporals}{\ensuremath{\mathbb{T}}}
\newcommand{\signals}{\ensuremath{\mathbb{S}}}
\newcommand{\pats}{\ensuremath{\mathbb{P}}}
\newcommand{\arbitrary}{\ensuremath{\mathbb{X}}}
\newcommand{\signalids}{\ensuremath{\Gamma_{\signals}}}
\newcommand{\busids}{\ensuremath{\Gamma_{\buses}}}
\newcommand{\enumids}{\ensuremath{\Gamma_{\enums}}}
\newcommand{\natids}{\ensuremath{\Gamma_{\nats}}}
\newcommand{\boolids}{\ensuremath{\Gamma_{\bools}}}
\newcommand{\temporalids}{\ensuremath{\Gamma_{\temporals}}}
\newcommand{\atypeids}{\ensuremath{\Gamma_{\mathcal{S}_{\arbitrary}}}}
\newcommand{\lstilde}{{\ensuremath{\color{codecolor} \raisebox{-9pt}{\scalebox{1.7}{\textasciitilde}}}}}
\newcommand{\tlsfsec}[1]{\ensuremath{\langle \text{\textit{#1}} \rangle}}
\newcommand{\tlsfid}[1]{\ensuremath{\langle \text{\textit{#1}} \rangle}}
\newcommand{\inputs}{\ensuremath{\mathcal{I}}}
\newcommand{\outputs}{\ensuremath{\mathcal{O}}}
\newcommand{\sep}{\ensuremath{\ \ | \ \ }}
\newcommand{\syntcomp}{SYNTCOMP\xspace}
\usepackage{tikz}

\usetikzlibrary{calc}
\usetikzlibrary{automata}
\usetikzlibrary{backgrounds}
\usetikzlibrary{decorations.pathreplacing}

\renewcommand{\subsubsection}[1]{\medskip \noindent {\bf #1.}}

\newcommand{\src}[1]{\texttt{\lstinline!#1!}}
\newcommand{\secref}[1]{Sect.~\ref{#1}}

\lstdefinestyle{TLSF}{
  belowcaptionskip=1\baselineskip,
  breaklines=true,
  frame=L,
  xleftmargin=\parindent,
  language=C,
  showstringspaces=false,
  basicstyle=\ttfamily\color{codecolor},
  keywordstyle=\ttfamily\color{codecolor},
  commentstyle=\ttfamily\color{codecolor},
  identifierstyle=\text\ttfamily\color{codecolor},
  stringstyle=\ttfamily\color{codecolor},
}

\lstdefinestyle{tlsf_keyword}{
  belowcaptionskip=1\baselineskip,
  breaklines=true,
  frame=L,
  xleftmargin=\parindent,
  language=C,
  showstringspaces=false,
  basicstyle=\ttfamily\color{codecolor},
  keywordstyle=\ttfamily\color{codecolor},
  commentstyle=\ttfamily\color{codecolor},
  identifierstyle=\text\ttfamily\color{codecolor},
  stringstyle=\ttfamily\color{codecolor},
}

\lstdefinestyle{tlsf_enumtyp}{
  belowcaptionskip=1\baselineskip,
  breaklines=true,
  frame=L,
  xleftmargin=\parindent,
  language=C,
  showstringspaces=false,
  basicstyle=\ttfamily\color{violet!90!black},
  keywordstyle=\ttfamily\color{violet!90!black},
  commentstyle=\ttfamily\color{violet!90!black},
  identifierstyle=\text\ttfamily\color{violet!90!black},
  stringstyle=\ttfamily\color{violet!90!black},
}

\lstdefinestyle{tlsf_enumval}{
  belowcaptionskip=1\baselineskip,
  breaklines=true,
  frame=L,
  xleftmargin=\parindent,
  language=C,
  showstringspaces=false,
  basicstyle=\ttfamily\color{red!60!black},
  keywordstyle=\ttfamily\color{red!60!black},
  commentstyle=\ttfamily\color{red!60!black},
  identifierstyle=\text\ttfamily\color{red!60!black},
  stringstyle=\ttfamily\color{red!60!black},
}

\lstdefinestyle{tlsf_semanti}{
  belowcaptionskip=1\baselineskip,
  breaklines=true,
  frame=L,
  xleftmargin=\parindent,
  language=C,
  showstringspaces=false,
  basicstyle=\ttfamily\color{gray!70!black},
  keywordstyle=\ttfamily\color{gray!70!black},
  commentstyle=\ttfamily\color{gray!70!black},
  identifierstyle=\text\ttfamily\color{gray!70!black},
  stringstyle=\ttfamily\color{gray!70!black},
}

\lstdefinestyle{tlsf_istring}{
  belowcaptionskip=1\baselineskip,
  breaklines=true,
  frame=L,
  xleftmargin=\parindent,
  language=C,
  showstringspaces=false,
  basicstyle=\ttfamily\color{red!30!orange!80!black},
  keywordstyle=\ttfamily\color{red!30!orange!80!black},
  commentstyle=\ttfamily\color{red!30!orange!80!black},
  identifierstyle=\text\ttfamily\color{red!30!orange!80!black},
  stringstyle=\ttfamily\color{red!30!orange!80!black},
}

\lstdefinestyle{tlsf_variabl}{
  belowcaptionskip=1\baselineskip,
  breaklines=true,
  frame=L,
  xleftmargin=\parindent,
  language=C,
  showstringspaces=false,
  basicstyle=\ttfamily\color{green!50!black},
  keywordstyle=\ttfamily\color{green!50!black},
  commentstyle=\ttfamily\color{green!50!black},
  identifierstyle=\text\ttfamily\color{green!50!black},
  stringstyle=\ttfamily\color{green!50!black},
}

\lstdefinestyle{tlsf_operato}{
  belowcaptionskip=1\baselineskip,
  breaklines=true,
  frame=L,
  xleftmargin=\parindent,
  language=C,
  showstringspaces=false,
  basicstyle=\ttfamily\color{cyan!50!black},
  keywordstyle=\ttfamily\color{cyan!50!black},
  commentstyle=\ttfamily\color{cyan!50!black},
  identifierstyle=\text\ttfamily\color{cyan!50!black},
  stringstyle=\ttfamily\color{cyan!50!black},
}

\lstdefinestyle{tlsf_default}{
  belowcaptionskip=1\baselineskip,
  breaklines=true,
  frame=L,
  xleftmargin=\parindent,
  language=C,
  showstringspaces=false,
  basicstyle=\ttfamily\color{black},
  keywordstyle=\ttfamily\color{black},
  commentstyle=\ttfamily\color{black},
  identifierstyle=\text\ttfamily\color{black},
  stringstyle=\ttfamily\color{black},
}

\lstdefinestyle{tlsf_comment}{
  belowcaptionskip=1\baselineskip,
  breaklines=true,
  frame=L,
  xleftmargin=\parindent,
  language=C,
  showstringspaces=false,
  basicstyle=\ttfamily\color{orange!60!black},
  keywordstyle=\ttfamily\color{orange!60!black},
  commentstyle=\ttfamily\color{orange!60!black},
  identifierstyle=\text\ttfamily\color{orange!60!black},
  stringstyle=\ttfamily\color{orange!60!black},
}

\DeclareMathAlphabet{\mathcal}{OMS}{cmsy}{m}{n}

\providecommand{\thisvolume}[1]{this volume of {\sl Electronic
  Proceedings in Theoretical Computer Science}. Open Publishing Association}

\title{A High-Level LTL Synthesis Format: \\ TLSF v1.1}
\author{Swen Jacobs
\institute{Saarland University\\ Saarbr\"ucken, Germany}
\email{jacobs@react.uni-saarland.de}
\and
Felix Klein
\institute{Saarland University\\ Saarbr\"ucken, Germany}
\email{klein@react.uni-saarland.de}
\and
Sebastian Schirmer
\institute{Saarland University\\ Saarbr\"ucken, Germany}
\email{s9sescir@stud.uni-saarland.de}
}
\def\titlerunning{A High-Level LTL Synthesis Format}
\def\authorrunning{S. Jacobs, F. Klein \& S. Schirmer}
\begin{document}
\newcommand{\arbiter}[2]{
  \begin{scope}[shift={(#1,#2)}]
    \def\h{4}
    \def\w{5}

    \coordinate (ANEW) at (0.5*\w-\w/2,\h/2);
    \coordinate (ASIG) at (0.3*\w-\w/2,-\h/2);
    \coordinate (AREADY) at (0.7*\w-\w/2,-\h/2);
    \coordinate (AR0) at (-\w/2,0.75*\h-\h/2);
    \coordinate (AR1) at (-\w/2,0.6*\h-\h/2);
    \coordinate (ARd) at (-\w/2,0.475*\h-\h/2);
    \coordinate (ARn) at (-\w/2,0.3*\h-\h/2);
    \coordinate (AG0) at (\w/2,0.75*\h-\h/2);
    \coordinate (AG1) at (\w/2,0.6*\h-\h/2);
    \coordinate (AGd) at (\w/2,0.475*\h-\h/2);
    \coordinate (AGn) at (\w/2,0.3*\h-\h/2);
    \coordinate (AHGRANT) at (\w/2+1.5*\a,0);
    \coordinate (AHBUSREQ) at (-\w/2-1.5*\a,0);

    \draw[fill=cmodule] (-\w/2,-\h/2) rectangle (\w/2,\h/2);
    \node at (0,-0.05*\h) {\large\bf ARBITER};
    
    \path[thick]
    (ANEW) edge[<-] +(0,-\a)
    (ASIG) edge[<-] +(0,\a)
    (AREADY) edge[->] +(0,\a)
    (AR0) edge[->] +(\a,0)
    (AR1) edge[->] +(\a,0)
    (ARn) edge[->] +(\a,0)
    (AG0) edge[<-] +(-\a,0)
    (AG1) edge[<-] +(-\a,0)
    (AGn) edge[<-] +(-\a,0)
    ;
    
    \path
    (AG0) edge  +(\a/2,0)
    (AG1) edge  +(\a/2,0)
    (AGn) edge  +(\a/2,0)
    ($ (AG0) + (\a/2,0) $) edge ([yshift=4] AHGRANT)
    ($ (AG1) + (\a/2,0) $) edge ([yshift=2] AHGRANT)
    ($ (AGn) + (\a/2,0) $) edge ([yshift=-4] AHGRANT)
    ;

    \path
    (AR0) edge  +(-\a/2,0)
    (AR1) edge  +(-\a/2,0)
    (ARn) edge  +(-\a/2,0)
    ($ (AR0) + (-\a/2,0) $) edge ([yshift=4] AHBUSREQ)
    ($ (AR1) + (-\a/2,0) $) edge ([yshift=2] AHBUSREQ)
    ($ (ARn) + (-\a/2,0) $) edge ([yshift=-4] AHBUSREQ)
    ;

    \node at ($ (ANEW) + (0,-1.5*\a) $) {\small \textsc{decide}};
    \node at ($ (ASIG) + (0,1.5*\a) $) {\small \textsc{busreq}};
    \node at ($ (AREADY) + (0,1.5*\a) $) {\small \textsc{allready}};

    \node[anchor=west] at ($ (AR0) + (0.8*\a,0) $) {\small \textsc{hbusreq}$ _{0} $};        
    \node[anchor=west] at ($ (AR1) + (0.8*\a,0) $) {\small \textsc{hbusreq}$ _{1} $};        
    \node[anchor=west] at ($ (ARd) + (0.5*\a,0) $) {$ \vdots $};        
    \node[anchor=west] at ($ (ARn) + (0.8*\a,0) $) {\small \textsc{hbusreq}$ _{n-1} $};        

    \node[anchor=east] at ($ (AG0) + (-0.8*\a,0) $) {\small \textsc{hgrant}$ _{0} $};        
    \node[anchor=east] at ($ (AG1) + (-0.8*\a,0) $) {\small \textsc{hgrant}$ _{1} $};        
    \node[anchor=east] at ($ (AGd) + (-0.5*\a,0) $) {$ \vdots $};        
    \node[anchor=east] at ($ (AGn) + (-0.8*\a,0) $) {\small \textsc{hgrant}$ _{n-1} $};        
  \end{scope}
}

\newcommand{\lock}[2]{
  \begin{scope}[shift={(#1,#2)}]
    \def\h{4}
    \def\w{4.5}

    \coordinate (MSIG) at (0.5*\w-\w/2,\h/2);
    \coordinate (MUPD) at (0.5*\w-\w/2,-\h/2); 
    \coordinate (MX0) at (-\w/2,0.75*\h-\h/2);
    \coordinate (MX1) at (-\w/2,0.6*\h-\h/2);
    \coordinate (MXd) at (-\w/2,0.475*\h-\h/2);
    \coordinate (MXn) at (-\w/2,0.3*\h-\h/2);
    \coordinate (ME0) at (\w/2,0.75*\h-\h/2);
    \coordinate (ME1) at (\w/2,0.6*\h-\h/2);
    \coordinate (MEd) at (\w/2,0.475*\h-\h/2);
    \coordinate (MEn) at (\w/2,0.3*\h-\h/2);
    \coordinate (MHGRANT) at (\w/2+1.5*\a,0);
    \coordinate (MHLOCK) at (-\w/2-1.5*\a,0);

    \draw[fill=cmodule] (-\w/2,-\h/2) rectangle (\w/2,\h/2);
    \node at (0,-0.05*\h) {\large\bf LOCK};
    
    \path[thick]
    (MSIG) edge[<-] +(0,-\a)
    (MUPD) edge[->] +(0,\a)
    (MX0) edge[->] +(\a,0)
    (MX1) edge[->] +(\a,0)
    (MXn) edge[->] +(\a,0)
    (ME0) edge[->] +(-\a,0)
    (ME1) edge[->] +(-\a,0)
    (MEn) edge[->] +(-\a,0)
    ;

    \path
    (ME0) edge  +(\a/2,0)
    (ME1) edge  +(\a/2,0)
    (MEn) edge  +(\a/2,0)
    ($ (ME0) + (\a/2,0) $) edge ([yshift=4] MHGRANT)
    ($ (ME1) + (\a/2,0) $) edge ([yshift=2] MHGRANT)
    ($ (MEn) + (\a/2,0) $) edge ([yshift=-4] MHGRANT)
    ;

    \path
    (MX0) edge  +(-\a/2,0)
    (MX1) edge  +(-\a/2,0)
    (MXn) edge  +(-\a/2,0)
    ($ (MX0) + (-\a/2,0) $) edge ([yshift=4] MHLOCK)
    ($ (MX1) + (-\a/2,0) $) edge ([yshift=2] MHLOCK)
    ($ (MXn) + (-\a/2,0) $) edge ([yshift=-4] MHLOCK)
    ;

    \node at ($ (MSIG) + (0,-1.5*\a) $) {\small \textsc{locked}};
    \node at ($ (MUPD) + (0,1.5*\a) $) {\small \textsc{decide}};

    \node[anchor=west] at ($ (MX0) + (0.8*\a,0) $) {\small \textsc{hlock}$ _{0} $};        
    \node[anchor=west] at ($ (MX1) + (0.8*\a,0) $) {\small \textsc{hlock}$ _{1} $};        
    \node[anchor=west] at ($ (MXd) + (0.5*\a,0) $) {$ \vdots $};        
    \node[anchor=west] at ($ (MXn) + (0.8*\a,0) $) {\small \textsc{hlock}$ _{n-1} $};        

    \node[anchor=east] at ($ (ME0) + (-0.8*\a,0) $) {\small \textsc{hgrant}$ _{0} $};        
    \node[anchor=east] at ($ (ME1) + (-0.8*\a,0) $) {\small \textsc{hgrant}$ _{1} $};        
    \node[anchor=east] at ($ (MEd) + (-0.5*\a,0) $) {$ \vdots $};        
    \node[anchor=east] at ($ (MEn) + (-0.8*\a,0) $) {\small \textsc{hgrant}$ _{n-1} $};        
  \end{scope}
}

\newcommand{\decode}[2]{
  \begin{scope}[shift={(#1,#2)}]
    \def\h{4}
    \def\w{3.6}

    \coordinate (INCR) at (\w/2,0.75*\h-\h/2);
    \coordinate (BURST4) at (\w/2,0.5*\h-\h/2);
    \coordinate (SINGLE) at (\w/2,0.25*\h-\h/2);
    \coordinate (HBURST) at (-\w/2,0.35*\h-\h/2);

    \draw[fill=cmodule] (-\w/2,-\h/2) rectangle (\w/2,\h/2);
    \node at (-0.15*\w,0.25*\h) {\large\bf DECODE};
    
    \path[thick]
    (INCR) edge[<-] +(-\a,0)
    (BURST4) edge[<-] +(-\a,0)
    (SINGLE) edge[<-] +(-\a,0)
    (HBURST) edge +(1.4*\a,0)
    ([yshift=2] HBURST) edge +(0.9*\a,0)
    ([yshift=4] HBURST) edge +(0.7*\a,0)
    ([yshift=-2] HBURST) edge +(0.9*\a,0)
    ([yshift=-4] HBURST) edge +(0.7*\a,0)
    ($ (HBURST) + (0.6*\a,-0.3) $) edge[bend left=40,ultra thick,cap=round] ($ (HBURST) + (1.4*\a,0) $)
    ($ (HBURST) + (0.6*\a,0.3) $) edge[bend right=40,ultra thick,cap=round] ($ (HBURST) + (1.4*\a,0) $)
    ;

    \node[anchor=east] at ($ (INCR) + (-0.8*\a,0) $) {\small \textsc{incr}};
    \node[anchor=east] at ($ (BURST4) + (-0.8*\a,0) $) {\small \textsc{burst}\scalebox{0.85}{4}};
    \node[anchor=east] at ($ (SINGLE) + (-0.8*\a,0) $) {\small \textsc{single}};
    \node[anchor=west] at ($ (HBURST) + (1.4*\a,0) $) {\small \textsc{hburst}};
  \end{scope}
}

\newcommand{\encode}[2]{
  \begin{scope}[shift={(#1,#2)}]
    \def\h{4}
    \def\w{4}

    \coordinate (HMASTER) at (-\w/2,0.65*\h-\h/2);
    \coordinate (HMASTERS) at ($ (HMASTER) + (\w - 1.5*\a,0) $);
    \coordinate (EMASTER) at ($ (HMASTER) + (\w,0) $);
    \coordinate (EHREADY) at (-\w/2,0.65*\h-\h/2);
    \coordinate (EG0) at (0.25*\w-\w/2,-\h/2);
    \coordinate (EG1) at (0.4*\w-\w/2,-\h/2);
    \coordinate (EGd) at (0.525*\w-\w/2,-\h/2);
    \coordinate (EGn) at (0.7*\w-\w/2,-\h/2);
    \coordinate (DHGRANT) at (0,-\h/2-1.5*\a);

    \draw[fill=cmodule] (-\w/2,-\h/2) rectangle (\w/2,\h/2);
    \node at (0,0.35*\h) {\large\bf ENCODE};

    \path
    (EG0) edge +(0,-\a/2)
    (EG1) edge +(0,-\a/2)
    (EGn) edge +(0,-\a/2)
    ($ (EG0) - (0,\a/2) $) edge ([xshift=-4] DHGRANT)
    ($ (EG1) - (0,\a/2) $) edge ([xshift=-2] DHGRANT)
    ($ (EGn) - (0,\a/2) $) edge ([xshift=4] DHGRANT)
    ;      
    
    \path[thick]
    (HMASTERS) edge +(1.4*\a,0)
    (EHREADY) edge[->] +(\a,0)
    (EG0) edge[->] +(0,\a)
    (EG1) edge[->] +(0,\a)
    (EGn) edge[->] +(0,\a)
    ([yshift=2] HMASTERS) edge +(0.9*\a,0)
    ([yshift=4] HMASTERS) edge +(0.7*\a,0)
    ([yshift=-2] HMASTERS) edge +(0.9*\a,0)
    ([yshift=-4] HMASTERS) edge +(0.7*\a,0)
    ($ (HMASTERS) + (0.6*\a,-0.3) $) edge[bend left=40,ultra thick,cap=round] ($ (HMASTERS) + (1.4*\a,0) $)
    ($ (HMASTERS) + (0.6*\a,0.3) $) edge[bend right=40,ultra thick,cap=round] ($ (HMASTERS) + (1.4*\a,0) $)
    ;

    \node[anchor=east] at (HMASTERS) {\small \textsc{hmaster}};
    \node[anchor=west] at ($ (EHREADY) + (\a,0) $) {\small \textsc{hready}};
    \node[anchor=west,rotate=90] at ($ (EG0) + (0,0.8*\a) $) {\small \textsc{hgrant}$ _{0} $};
    \node[anchor=west,rotate=90] at ($ (EG1) + (0,0.8*\a) $) {\small \textsc{hgrant}$ _{1} $};
    \node[anchor=west,rotate=90] at ($ (EGd) + (0,0.5*\a) $) {\small $ \vdots $};
    \node[anchor=west,rotate=90] at ($ (EGn) + (0,0.8*\a) $) {\small \textsc{hgrant}$ _{n-1} $};
  \end{scope}
}

\newcommand{\shift}[2]{
  \begin{scope}[shift={(#1,#2)}]
    \def\h{2}
    \def\w{4}

    \coordinate (HMASTLOCK) at (\w/2,0.35*\h-\h/2);
    \coordinate (FHREADY) at (-\w/2,0.2*\h-\h/2); 
    \coordinate (SLOCKED) at (-\w/2,0.5*\h-\h/2); 

    \draw[fill=cmodule] (-\w/2,-\h/2) rectangle (\w/2,\h/2);
    \node at (0,0.28*\h) {\large\bf SHIFT};
    
    \path[thick]
    (HMASTLOCK) edge[<-] +(-\a,0)
    (FHREADY) edge[->] +(\a,0)
    (SLOCKED) edge[->] +(\a,0)
    ;

    \node[anchor=east] at ($ (HMASTLOCK) + (-\a,0) $) {\small \textsc{hmastlock}};
    \node[anchor=west] at ($ (FHREADY) + (\a,0) $) {\small \textsc{hready}};
    \node[anchor=west] at ($ (SLOCKED) + (\a,0) $) {\small \textsc{locked}};
  \end{scope}
}

\newcommand{\tincr}[2]{
  \begin{scope}[shift={(#1,#2)}]
    \def\h{4}
    \def\w{4.5}

    \coordinate (IREADY) at (0.5*\w-\w/2,\h/2);
    \coordinate (IHREADY) at (0.2*\w-\w/2,-\h/2);
    \coordinate (ILOCK) at (0.4*\w-\w/2,-\h/2);
    \coordinate (IDECIDE) at (0.6*\w-\w/2,-\h/2);
    \coordinate (ISIG) at (0.8*\w-\w/2,-\h/2);
    \coordinate (IBUSREQ) at (-\w/2,0.7*\h-\h/2);

    \draw[fill=cmodule] (-\w/2,-\h/2) rectangle (\w/2,\h/2);
    \node at (0.1*\w,0.1*\h) {\large\bf TINCR};
    
    \path[thick]
    (IREADY) edge[<-] +(0,-\a)      
    (IHREADY) edge[->] +(0,\a)
    (ILOCK) edge[->] +(0,\a)
    (IDECIDE) edge[->] +(0,\a)
    (ISIG) edge[->] +(0,\a)
    (IBUSREQ) edge[->] +(\a,0)
    ;

    \node at ($ (IREADY) + (0,-1.5*\a) $) {\small \textsc{ready}$ _{1} $};
    \node[rotate=90,anchor=west] at ($ (IHREADY) + (0,\a) $) {\small \textsc{hready}};
    \node[rotate=90,anchor=west] at ($ (ILOCK) + (0,\a) $) {\small \textsc{locked}};
    \node[rotate=90,anchor=west] at ($ (IDECIDE) + (0,\a) $) {\small \textsc{decide}};
    \node[rotate=90,anchor=west] at ($ (ISIG) + (0,\a) $) {\small \textsc{incr}};
    \node[anchor=west] at ($ (IBUSREQ) + (\a,0) $) {\small \textsc{busreq}};
  \end{scope}
}

\newcommand{\tburstfour}[2]{
  \begin{scope}[shift={(#1,#2)}]
    \def\h{4}
    \def\w{4.5}

    \coordinate (BREADY) at (0.5*\w-\w/2,\h/2);
    \coordinate (BHREADY) at (0.2*\w-\w/2,-\h/2);
    \coordinate (BLOCK) at (0.4*\w-\w/2,-\h/2);
    \coordinate (BDECIDE) at (0.6*\w-\w/2,-\h/2);
    \coordinate (BSIG) at (0.8*\w-\w/2,-\h/2);

    \draw[fill=cmodule] (-\w/2,-\h/2) rectangle (\w/2,\h/2);
    \node at (0,0.1*\h) {\large\bf TBURST4};
    
    \path[thick]
    (BREADY) edge[<-] +(0,-\a)      
    (BHREADY) edge[->] +(0,\a)
    (BLOCK) edge[->] +(0,\a)
    (BDECIDE) edge[->] +(0,\a)
    (BSIG) edge[->] +(0,\a)
    ;

    \node at ($ (BREADY) + (0,-1.5*\a) $) {\small \textsc{ready}$ _{2} $};
    \node[rotate=90,anchor=west] at ($ (BHREADY) + (0,\a) $) {\small \textsc{hready}};
    \node[rotate=90,anchor=west] at ($ (BLOCK) + (0,\a) $) {\small \textsc{locked}};
    \node[rotate=90,anchor=west] at ($ (BDECIDE) + (0,\a) $) {\small \textsc{decide}};
    \node[rotate=90,anchor=west] at ($ (BSIG) + (0,\a) $) {\small \textsc{burst}\scalebox{0.85}{4}};
  \end{scope}
}

\newcommand{\tsingle}[2]{
  \begin{scope}[shift={(#1,#2)}]
    \def\h{4}
    \def\w{4}

    \coordinate (SREADY) at (0.5*\w-\w/2,\h/2);
    \coordinate (SHREADY) at (0.2*\w-\w/2,-\h/2);
    \coordinate (SLOCK) at (0.4*\w-\w/2,-\h/2);
    \coordinate (SDECIDE) at (0.6*\w-\w/2,-\h/2);
    \coordinate (SSIG) at (0.8*\w-\w/2,-\h/2);

    \draw[fill=cmodule] (-\w/2,-\h/2) rectangle (\w/2,\h/2);
    \node at (0,0.1*\h) {\large\bf TSINGLE};
    
    \path[thick]
    (SREADY) edge[<-] +(0,-\a)      
    (SHREADY) edge[->] +(0,\a)
    (SLOCK) edge[->] +(0,\a)
    (SDECIDE) edge[->] +(0,\a)
    (SSIG) edge[->] +(0,\a)
    ;

    \node at ($ (SREADY) + (0,-1.5*\a) $) {\small \textsc{ready}$ _{3} $};
    \node[rotate=90,anchor=west] at ($ (SHREADY) + (0,\a) $) {\small \textsc{hready}};
    \node[rotate=90,anchor=west] at ($ (SLOCK) + (0,\a) $) {\small \textsc{locked}};
    \node[rotate=90,anchor=west] at ($ (SDECIDE) + (0,\a) $) {\small \textsc{decide}};
    \node[rotate=90,anchor=west] at ($ (SSIG) + (0,\a) $) {\small \textsc{single}};
  \end{scope}
}

\newcommand{\mand}[2]{
  \begin{scope}[shift={(#1,#2)}]
    \def\h{1.2}
    \def\w{2}

    \coordinate (R1) at (-0.3,-\h/2);
    \coordinate (R2) at (0,-\h/2);
    \coordinate (R3) at (0.3,-\h/2);
    \coordinate (AA) at (0,\h/2);

    \draw[fill=cmodule] (-\w/2,-\h/2) rectangle (\w/2,\h/2);
    \node at (0,0) {\large\bf AND};

    \path[thick]
    (R1) edge[->] +(0,\a)      
    (R2) edge[->] +(0,\a)      
    (R3) edge[->] +(0,\a)      
    (AA) edge[<-] +(0,-\a)      
    ;
  \end{scope}
}

\newsavebox{\architecture}
\begin{lrbox}{\architecture}
  \begin{tikzpicture}
    \def\a{0.3}
    \definecolor{cmodule}{gray}{0.9}
    \def\dd{0.7}

    \def\fleft{-9.5}
    \def\fright{7.6}
    \def\ftop{16.9}
    \def\fbot{-8.5}

    \def\cx{1.5}
    \def\cy{5.3}

    \arbiter{-4.8}{8}
    \lock{-4.8}{13.8}
    \decode{-7.1}{-6}
    \encode{5}{11.5}
    \shift{5}{15.4}
    \tincr{-6}{0}
    \tburstfour{-1}{0}
    \tsingle{3.75}{0}
    \mand{-1}{4}

    \draw (ANEW) -- (MUPD);
    \draw (ASIG) |- (\fleft+0.6,\cy) |- (IBUSREQ);
    \draw (INCR) -| (ISIG);
    \draw (BURST4) -| (BSIG);
    \draw (SINGLE) -| (SSIG);
    \draw (SHREADY) -- +(0,-1.3) node[fill,circle,inner sep=1pt] {} -| ($ (SREADY) + (3.3,2.3) $);
    \draw (SDECIDE) -- +(0,-0.7) node[fill,circle,inner sep=1pt] {}-| ($ (SREADY) + (2.7,1.7) $);
    \draw (SLOCK) -- +(0,-1)  node[fill,circle,inner sep=1pt] {} -| ($ (SREADY) + (3,2) $);
    \draw ($ (SREADY) + (3,2) $) -| (\cx,\ftop-0.5); 
    \draw ($ (SREADY) + (2.7,1.7) $) -| (\cx-0.3,5); 
    \draw ($ (SREADY) + (3.3,2.3) $) -| (\cx+0.3,5); 
    \draw (FHREADY) -| (\cx+0.3,5); 
    \draw (MSIG) |- (\cx,\ftop-0.5); 
    \draw (\cx-0.3,5) |- ($ (ANEW) + (0,0.9) $) node[fill,circle,inner sep=1pt] {};
    \draw (SLOCKED) -- (SLOCKED -| \cx,0) node[fill,circle,inner sep=1pt] {};
    \draw (EHREADY) -- (EHREADY -| \cx+0.3,0) node[fill,circle,inner sep=1pt] {};
    \draw (IDECIDE) -- +(0,-0.7) -- +(9.65,-0.7);
    \draw (BDECIDE) -- +(0,-0.7) node[fill,circle,inner sep=1pt] {};
    \draw (ILOCK) -- +(0,-1) -- +(9.8,-1);
    \draw (BLOCK) -- +(0,-1) node[fill,circle,inner sep=1pt] {};
    \draw (IHREADY) -- +(0,-1.3)  node[fill,circle,inner sep=1pt] {} -- +(9.95,-1.3); 
    \draw (BHREADY) -- +(0,-1.3) node[fill,circle,inner sep=1pt] {};

    \foreach \i in {-4,-2,0,2,4} 
      \draw ([xshift=\i] DHGRANT) -- ([xshift=\i,yshift=\i] AHGRANT -| DHGRANT)  node[fill,circle,inner sep=0.6pt] {};

    \foreach \i in {-4,-2,0,2,4} 
      \draw ([yshift=\i] MHGRANT) -| ($ ([xshift=\i,yshift=\i] AHGRANT) + (1.5,0) $) node[fill,circle,inner sep=0.6pt] {};

    \draw (BREADY) -- (R2); 
    \draw (IREADY) -- +(0,0.7) -| (R1);
    \draw (SREADY) -- +(0,0.7) -| (R3);

    \draw (AA) -- (AA |- 0,\cy) -| (AREADY);

    \node[above,yshift=5,xshift=6] at (AHGRANT -| \fright+\dd,0) {\small \textsc{HGRANT}};
    \foreach \i in {-4,-2,0,2,4} 
      \draw ([yshift=\i] AHGRANT) -- ([yshift=\i] AHGRANT -| \fright+\dd,0);

    \node[above,yshift=5,xshift=-3] at (MHLOCK -| \fleft-\dd,0) {\small \textsc{HLOCK}};
    \foreach \i in {-4,-2,0,2,4} 
      \draw ([yshift=\i] MHLOCK) -- ([yshift=\i] MHLOCK -| \fleft-\dd,0);

    \node[above,yshift=5,xshift=-9] at (AHBUSREQ -| \fleft-\dd,0) {\small \textsc{HBUSREQ}};
    \foreach \i in {-4,-2,0,2,4} 
      \draw ([yshift=\i] AHBUSREQ) -- ([yshift=\i] AHBUSREQ -| \fleft-\dd,0);

    \draw (HMASTLOCK) -- (HMASTLOCK -| \fright+\dd,0) node[above,xshift=16,yshift=1] {\small \textsc{HMASTLOCK}};

    \node[above,yshift=5,xshift=9.5] at (HMASTER -| \fright+\dd,0) {\small \textsc{HMASTER}};
    \foreach \i in {-4,-2,0,2,4} 
      \draw ([yshift=\i] EMASTER) -- ([yshift=\i] EMASTER -| \fright+\dd,0);

    \node[above,yshift=5,xshift=-5] at (HBURST -| \fleft-\dd,0) {\small \textsc{HBURST}};
    \foreach \i in {-4,-2,0,2,4} 
      \draw ([yshift=\i] HBURST) -- ([yshift=\i] HBURST -| \fleft-\dd,0);

    \coordinate (P1) at ($ (IHREADY) + (0,-1.3) $);
    \draw (P1) -- (P1 -| \fleft-\dd,0) node[above,xshift=-5,yshift=1] {\small \textsc{HREADY}};

    \draw[ultra thick] (\fleft,\fbot) rectangle (\fright,\ftop);
  \end{tikzpicture}
\end{lrbox}

\newsavebox{\cPArbiter}
\begin{lrbox}{\cPArbiter}
\begin{minipage}{0.7\textwidth}
\scriptsize
\noindent
\lstinline[style=tlsf_keyword]~INFO~%
\lstinline[style=tlsf_default]~ {~\\%
\lstinline[style=tlsf_keyword]~  TITLE~%
\lstinline[style=tlsf_default]~:       ~%
\lstinline[style=tlsf_istring]~"A Parameterized Arbiter"~\\%
\lstinline[style=tlsf_keyword]~  DESCRIPTION~%
\lstinline[style=tlsf_default]~: ~%
\lstinline[style=tlsf_istring]~"An arbiter, parameterized in the number of clients"~\\%
\lstinline[style=tlsf_keyword]~  SEMANTICS~%
\lstinline[style=tlsf_default]~:   ~%
\lstinline[style=tlsf_semanti]~Mealy~\\%
\lstinline[style=tlsf_keyword]~  TARGET~%
\lstinline[style=tlsf_default]~:      ~%
\lstinline[style=tlsf_semanti]~Mealy~\\%
\lstinline[style=tlsf_default]~}~\\%
\lstinline[style=tlsf_keyword]~GLOBAL ~%
\lstinline[style=tlsf_default]~{~\\%
\lstinline[style=tlsf_keyword]~  PARAMETERS ~%
\lstinline[style=tlsf_default]~{~\\%
\lstinline[style=tlsf_comment]~    // two clients~\\%
\lstinline[style=tlsf_variabl]~    n~%
\lstinline[style=tlsf_default]~ = 2;~\\%
\lstinline[style=tlsf_default]~  }~\\%
\lstinline[style=tlsf_keyword]~  DEFINITIONS~%
\lstinline[style=tlsf_default]~ {~\\%
\lstinline[style=tlsf_comment]~    // mutual exclusion~\\%
\lstinline[style=tlsf_variabl]~    mutual~%
\lstinline[style=tlsf_default]~(b) =~\\%
\lstinline[style=tlsf_operato]~      ||~%
\lstinline[style=tlsf_default]~[i ~%
\lstinline[style=tlsf_operato]~IN~%
\lstinline[style=tlsf_default]~ {0, 1 .. (~%
\lstinline[style=tlsf_operato]~SIZEOF ~%
\lstinline[style=tlsf_default]~b)~%
\lstinline[style=tlsf_operato]~ - ~%
\lstinline[style=tlsf_default]~1}]~\\%
\lstinline[style=tlsf_operato]~        &&~%
\lstinline[style=tlsf_default]~[j ~%
\lstinline[style=tlsf_operato]~IN~%
\lstinline[style=tlsf_default]~ {0, 1 .. (~%
\lstinline[style=tlsf_operato]~SIZEOF ~%
\lstinline[style=tlsf_default]~b)~%
\lstinline[style=tlsf_operato]~ - ~%
\lstinline[style=tlsf_default]~1} ~%
\lstinline[style=tlsf_operato]~(\)~%
\lstinline[style=tlsf_default]~ {i}]~\\%
\lstinline[style=tlsf_operato]~          !~%
\lstinline[style=tlsf_default]~(b[i] ~%
\lstinline[style=tlsf_operato]~&&~%
\lstinline[style=tlsf_default]~ b[j]);~\\%
\lstinline[style=tlsf_comment]~    // the Request-Response condition~\\%
\lstinline[style=tlsf_variabl]~    reqres~%
\lstinline[style=tlsf_default]~(req, res) =~\\%
\lstinline[style=tlsf_operato]~      G~%
\lstinline[style=tlsf_default]~ (req ~%
\lstinline[style=tlsf_operato]~-> F~%
\lstinline[style=tlsf_default]~ res);~\\%
\lstinline[style=tlsf_default]~  }~\\%
\lstinline[style=tlsf_default]~}~\\%
\lstinline[style=tlsf_comment]~/* ensure mutual exclusion on the output bus and guarantee~\\%
\lstinline[style=tlsf_comment]~   that each request on the input bus is eventually granted */~\\%
\lstinline[style=tlsf_keyword]~MAIN~%
\lstinline[style=tlsf_default]~ {~\\%
\lstinline[style=tlsf_keyword]~  INPUTS~%
\lstinline[style=tlsf_default]~ {~\\%
\lstinline[style=tlsf_variabl]~    r~%
\lstinline[style=tlsf_default]~[n];~\\%
\lstinline[style=tlsf_default]~  }~\\%
\lstinline[style=tlsf_keyword]~  OUTPUTS~%
\lstinline[style=tlsf_default]~ {~\\%
\lstinline[style=tlsf_variabl]~    g~%
\lstinline[style=tlsf_default]~[n];~\\%
\lstinline[style=tlsf_default]~  }~\\%
\lstinline[style=tlsf_keyword]~  ASSERT~%
\lstinline[style=tlsf_default]~ {~\\%
\lstinline[style=tlsf_default]~    mutual(g);~\\%
\lstinline[style=tlsf_default]~  }~\\%
\lstinline[style=tlsf_keyword]~  GUARANTEE~%
\lstinline[style=tlsf_default]~ {~\\%
\lstinline[style=tlsf_operato]~    &&~%
\lstinline[style=tlsf_default]~[0 ~%
\lstinline[style=tlsf_operato]~<=~%
\lstinline[style=tlsf_default]~ i ~%
\lstinline[style=tlsf_operato]~<~%
\lstinline[style=tlsf_default]~ n]~\\%
\lstinline[style=tlsf_default]~      reqres(r[i], g[i]);~\\%
\lstinline[style=tlsf_default]~  }~\\
\lstinline[style=tlsf_default]~}~
\end{minipage}
\end{lrbox}

\newsavebox{\cDecodeA}
\begin{lrbox}{\cDecodeA}
\begin{minipage}{0.4\textwidth}
\scriptsize
\lstinline[style=tlsf_keyword]~INFO~%
\lstinline[style=tlsf_default]~ {~\\%
\lstinline[style=tlsf_keyword]~  TITLE~%
\lstinline[style=tlsf_default]~:       ~%
\lstinline[style=tlsf_istring]~"AMBA AHB Arbiter"~\\%
\lstinline[style=tlsf_keyword]~  DESCRIPTION~%
\lstinline[style=tlsf_default]~: ~%
\lstinline[style=tlsf_istring]~"Component: Decode"~\\%
\lstinline[style=tlsf_keyword]~  SEMANTICS~%
\lstinline[style=tlsf_default]~:   ~%
\lstinline[style=tlsf_semanti]~Mealy~\\%
\lstinline[style=tlsf_keyword]~  TARGET~%
\lstinline[style=tlsf_default]~:      ~%
\lstinline[style=tlsf_semanti]~Mealy~\\%
\lstinline[style=tlsf_default]~}~\\%
\lstinline[style=tlsf_default]~~\\%
\lstinline[style=tlsf_keyword]~GLOBAL ~%
\lstinline[style=tlsf_default]~{~\\%
\lstinline[style=tlsf_keyword]~  DEFINITIONS~%
\lstinline[style=tlsf_default]~ {~\\%
\lstinline[style=tlsf_keyword]~    enum~%
\lstinline[style=tlsf_enumtyp]~ hburst~%
\lstinline[style=tlsf_default]~ =~\\%
\lstinline[style=tlsf_enumval]~      Single~%
\lstinline[style=tlsf_default]~: 00~\\%
\lstinline[style=tlsf_enumval]~      Burst4~%
\lstinline[style=tlsf_default]~: 10~\\%
\lstinline[style=tlsf_enumval]~      Incr~%
\lstinline[style=tlsf_default]~:   01~\\%
\lstinline[style=tlsf_default]~  }~\\%
\lstinline[style=tlsf_default]~}~\\%
\lstinline[style=tlsf_default]~~
\end{minipage}
\end{lrbox}

\newsavebox{\cDecodeB}
\begin{lrbox}{\cDecodeB}
\begin{minipage}{0.528\textwidth}
\scriptsize
\lstinline[style=tlsf_keyword]~MAIN~%
\lstinline[style=tlsf_default]~ {~\\%
\lstinline[style=tlsf_keyword]~  INPUTS~%
\lstinline[style=tlsf_default]~ {~\\%
\lstinline[style=tlsf_enumtyp]~    hburst~%
\lstinline[style=tlsf_variabl]~ HBURST~%
\lstinline[style=tlsf_default]~;~\\%
\lstinline[style=tlsf_default]~  }~\\%
\lstinline[style=tlsf_keyword]~  OUTPUTS~%
\lstinline[style=tlsf_default]~ {~\\%
\lstinline[style=tlsf_variabl]~    SINGLE~%
\lstinline[style=tlsf_default]~;~\\%
\lstinline[style=tlsf_variabl]~    BURST4~%
\lstinline[style=tlsf_default]~;~\\%
\lstinline[style=tlsf_variabl]~    INCR~%
\lstinline[style=tlsf_default]~;~\\%
\lstinline[style=tlsf_default]~    }~\\%
\lstinline[style=tlsf_keyword]~  ASSERT~%
\lstinline[style=tlsf_default]~ {~\\%
\lstinline[style=tlsf_default]~    HBURST ~%
\lstinline[style=tlsf_operato]~==~%
\lstinline[style=tlsf_enumval]~ Single ~%
\lstinline[style=tlsf_operato]~->~%
\lstinline[style=tlsf_default]~ SINGLE;~\\%
\lstinline[style=tlsf_default]~    HBURST ~%
\lstinline[style=tlsf_operato]~==~%
\lstinline[style=tlsf_enumval]~ Burst4 ~%
\lstinline[style=tlsf_operato]~->~%
\lstinline[style=tlsf_default]~ BURST4;~\\%
\lstinline[style=tlsf_default]~    HBURST ~%
\lstinline[style=tlsf_operato]~==~%
\lstinline[style=tlsf_enumval]~ Incr ~%
\lstinline[style=tlsf_operato]~->~%
\lstinline[style=tlsf_default]~ INCR;~\\%
\lstinline[style=tlsf_operato]~    !~%
\lstinline[style=tlsf_default]~(SINGLE ~%
\lstinline[style=tlsf_operato]~&&~%
\lstinline[style=tlsf_default]~ (BURST4 ~%
\lstinline[style=tlsf_operato]~||~%
\lstinline[style=tlsf_default]~ INCR)) ~%
\lstinline[style=tlsf_operato]~&&~%
\lstinline[style=tlsf_operato]~ !~%
\lstinline[style=tlsf_default]~(BURST4 ~%
\lstinline[style=tlsf_operato]~&&~%
\lstinline[style=tlsf_default]~ INCR);~\\%
\lstinline[style=tlsf_default]~  }~\\%
\lstinline[style=tlsf_default]~}~
\end{minipage}
\end{lrbox}

\newsavebox{\cArbiter}
\begin{lrbox}{\cArbiter}
\begin{minipage}{.49\textwidth}
\scriptsize
\lstinline[style=tlsf_keyword]~INFO~%
\lstinline[style=tlsf_default]~ {~\\%
\lstinline[style=tlsf_keyword]~  TITLE~%
\lstinline[style=tlsf_default]~:       ~%
\lstinline[style=tlsf_istring]~"AMBA AHB Arbiter"~\\%
\lstinline[style=tlsf_keyword]~  DESCRIPTION~%
\lstinline[style=tlsf_default]~: ~%
\lstinline[style=tlsf_istring]~"Component: Arbiter"~\\%
\lstinline[style=tlsf_keyword]~  SEMANTICS~%
\lstinline[style=tlsf_default]~:   ~%
\lstinline[style=tlsf_semanti]~Mealy~\\%
\lstinline[style=tlsf_keyword]~  TARGET~%
\lstinline[style=tlsf_default]~:      ~%
\lstinline[style=tlsf_semanti]~Mealy~\\%
\lstinline[style=tlsf_default]~}~\\%
\lstinline[style=tlsf_keyword]~GLOBAL ~%
\lstinline[style=tlsf_default]~{~\\%
\lstinline[style=tlsf_keyword]~  PARAMETERS ~%
\lstinline[style=tlsf_default]~{~\\%
\lstinline[style=tlsf_variabl]~    n~%
\lstinline[style=tlsf_default]~ = 2;~\\%
\lstinline[style=tlsf_default]~  }~\\%
\lstinline[style=tlsf_keyword]~  DEFINITIONS~%
\lstinline[style=tlsf_default]~ {~\\%
\lstinline[style=tlsf_comment]~    // mutual exclusion~\\%
\lstinline[style=tlsf_variabl]~    mutual~%
\lstinline[style=tlsf_default]~(b) =~\\%
\lstinline[style=tlsf_operato]~      ||~%
\lstinline[style=tlsf_default]~[i ~%
\lstinline[style=tlsf_operato]~IN~%
\lstinline[style=tlsf_default]~ {0, 1 .. (~%
\lstinline[style=tlsf_operato]~SIZEOF ~%
\lstinline[style=tlsf_default]~b)~%
\lstinline[style=tlsf_operato]~ - ~%
\lstinline[style=tlsf_default]~1}]~\\%
\lstinline[style=tlsf_operato]~        &&~%
\lstinline[style=tlsf_default]~[j ~%
\lstinline[style=tlsf_operato]~IN~%
\lstinline[style=tlsf_default]~ {0, 1 .. (~%
\lstinline[style=tlsf_operato]~SIZEOF ~%
\lstinline[style=tlsf_default]~b)~%
\lstinline[style=tlsf_operato]~ - ~%
\lstinline[style=tlsf_default]~1} ~%
\lstinline[style=tlsf_operato]~(\)~%
\lstinline[style=tlsf_default]~ {i}]~\\%
\lstinline[style=tlsf_operato]~          !~%
\lstinline[style=tlsf_default]~(b[i] ~%
\lstinline[style=tlsf_operato]~&&~%
\lstinline[style=tlsf_default]~ b[j]);~\\%
\lstinline[style=tlsf_default]~  }~\\%
\lstinline[style=tlsf_default]~}~\\%
\lstinline[style=tlsf_keyword]~MAIN~%
\lstinline[style=tlsf_default]~ {~\\%
\lstinline[style=tlsf_keyword]~  INPUTS~%
\lstinline[style=tlsf_default]~ {~\\%
\lstinline[style=tlsf_variabl]~    HBUSREQ~%
\lstinline[style=tlsf_default]~[n];~\\%
\lstinline[style=tlsf_variabl]~    ALLREADY~%
\lstinline[style=tlsf_default]~;~\\%
\lstinline[style=tlsf_default]~  }~\\%
\lstinline[style=tlsf_keyword]~  OUTPUTS~%
\lstinline[style=tlsf_default]~ {~\\%
\lstinline[style=tlsf_variabl]~    HGRANT~%
\lstinline[style=tlsf_default]~[n];~\\%
\lstinline[style=tlsf_variabl]~    BUSREQ~%
\lstinline[style=tlsf_default]~;~\\%
\lstinline[style=tlsf_variabl]~    DECIDE~%
\lstinline[style=tlsf_default]~;~\\%
\lstinline[style=tlsf_default]~  }~\\%
\lstinline[style=tlsf_keyword]~  INITIALLY~%
\lstinline[style=tlsf_default]~ {~\\%
\lstinline[style=tlsf_comment]~    // the component is initially idle~\\%
\lstinline[style=tlsf_default]~    ALLREADY;~\\%
\lstinline[style=tlsf_default]~  }~\\%
\lstinline[style=tlsf_keyword]~  ASSUME~%
\lstinline[style=tlsf_default]~ {~\\%
\lstinline[style=tlsf_comment]~    // the component is not eventually disabled~\\%
\lstinline[style=tlsf_operato]~    G F~%
\lstinline[style=tlsf_default]~ ALLREADY;~\\%
\lstinline[style=tlsf_default]~  }~\\%
\lstinline[style=tlsf_keyword]~  ASSERT~%
\lstinline[style=tlsf_default]~ {~\\%
\lstinline[style=tlsf_comment]~    // always exactely one master is granted~\\%
\lstinline[style=tlsf_default]~    mutual(HGRANT) ~%
\lstinline[style=tlsf_operato]~&& ||~%
\lstinline[style=tlsf_default]~[0 ~%
\lstinline[style=tlsf_operato]~<=~%
\lstinline[style=tlsf_default]~ i ~%
\lstinline[style=tlsf_operato]~<~%
\lstinline[style=tlsf_default]~ n] HGRANT[i];~\\%
\lstinline[style=tlsf_comment]~    // if not ready, the grants stay unchanged~\\%
\lstinline[style=tlsf_operato]~    &&~%
\lstinline[style=tlsf_default]~[0 ~%
\lstinline[style=tlsf_operato]~<=~%
\lstinline[style=tlsf_default]~ i ~%
\lstinline[style=tlsf_operato]~<~%
\lstinline[style=tlsf_default]~ n]~\\%
\lstinline[style=tlsf_default]~      (~%
\lstinline[style=tlsf_operato]~!~%
\lstinline[style=tlsf_default]~ALLREADY ~%
\lstinline[style=tlsf_operato]~->~%
\lstinline[style=tlsf_default]~ (~%
\lstinline[style=tlsf_operato]~X~%
\lstinline[style=tlsf_default]~ HGRANT[i] ~%
\lstinline[style=tlsf_operato]~<->~%
\lstinline[style=tlsf_default]~ HGRANT[i]));~\\%
\lstinline[style=tlsf_comment]~    // every request is eventually granted~\\%
\lstinline[style=tlsf_operato]~    &&~%
\lstinline[style=tlsf_default]~[0 ~%
\lstinline[style=tlsf_operato]~<=~%
\lstinline[style=tlsf_default]~ i ~%
\lstinline[style=tlsf_operato]~<~%
\lstinline[style=tlsf_default]~ n]~\\%
\lstinline[style=tlsf_default]~      (HBUSREQ[i] ~%
\lstinline[style=tlsf_operato]~-> F~%
\lstinline[style=tlsf_default]~ (~%
\lstinline[style=tlsf_operato]~!~%
\lstinline[style=tlsf_default]~HBUSREQ[i] ~%
\lstinline[style=tlsf_operato]~||~%
\lstinline[style=tlsf_default]~ HGRANT[i]));~\\%
\lstinline[style=tlsf_comment]~    // the BUSREQ signal mirrors the HBUSREQ[i]~\\%
\lstinline[style=tlsf_comment]~    // signal of the currently granted master i~\\%
\lstinline[style=tlsf_operato]~    &&~%
\lstinline[style=tlsf_default]~[0 ~%
\lstinline[style=tlsf_operato]~<=~%
\lstinline[style=tlsf_default]~ i ~%
\lstinline[style=tlsf_operato]~<~%
\lstinline[style=tlsf_default]~ n]~\\%
\lstinline[style=tlsf_default]~      (HGRANT[i] ~%
\lstinline[style=tlsf_operato]~->~%
\lstinline[style=tlsf_default]~ (BUSREQ ~%
\lstinline[style=tlsf_operato]~<->~%
\lstinline[style=tlsf_default]~ HBUSREQ[i]));~\\
\lstinline[style=tlsf_comment]~    // taking decisions requires to be idle~\\%
\lstinline[style=tlsf_operato]~    !~%
\lstinline[style=tlsf_default]~ALLREADY ~%
\lstinline[style=tlsf_operato]~-> !~%
\lstinline[style=tlsf_default]~DECIDE;~\\%
\lstinline[style=tlsf_comment]~    // granting another master triggers a decision~\\%
\lstinline[style=tlsf_default]~    DECIDE ~%
\lstinline[style=tlsf_operato]~<-> ||~%
\lstinline[style=tlsf_default]~[0 ~%
\lstinline[style=tlsf_operato]~<=~%
\lstinline[style=tlsf_default]~ i ~%
\lstinline[style=tlsf_operato]~<~%
\lstinline[style=tlsf_default]~ n]~\\%
\lstinline[style=tlsf_operato]~      !~%
\lstinline[style=tlsf_default]~(~%
\lstinline[style=tlsf_operato]~X~%
\lstinline[style=tlsf_default]~ HGRANT[i] ~%
\lstinline[style=tlsf_operato]~<->~%
\lstinline[style=tlsf_default]~ HGRANT[i]);~\\%
\lstinline[style=tlsf_comment]~    // if there is no request, master 0 is granted~\\%
\lstinline[style=tlsf_default]~    (~%
\lstinline[style=tlsf_operato]~&&~%
\lstinline[style=tlsf_default]~[0 ~%
\lstinline[style=tlsf_operato]~<=~%
\lstinline[style=tlsf_default]~ i ~%
\lstinline[style=tlsf_operato]~<~%
\lstinline[style=tlsf_default]~ n] ~%
\lstinline[style=tlsf_operato]~!~%
\lstinline[style=tlsf_default]~HBUSREQ[i]) ~%
\lstinline[style=tlsf_operato]~&&~%
\lstinline[style=tlsf_default]~ DECIDE~\\%
\lstinline[style=tlsf_operato]~       -> X~%
\lstinline[style=tlsf_default]~ HGRANT[0];~\\%
\lstinline[style=tlsf_default]~  }~\\%
\lstinline[style=tlsf_default]~}~
\end{minipage}
\end{lrbox}

\newsavebox{\cEncode}
\begin{lrbox}{\cEncode}
\begin{minipage}{.49\textwidth}
\scriptsize
\lstinline[style=tlsf_keyword]~INFO~%
\lstinline[style=tlsf_default]~ {~\\%
\lstinline[style=tlsf_keyword]~  TITLE~%
\lstinline[style=tlsf_default]~:       ~%
\lstinline[style=tlsf_istring]~"AMBA AHB Arbiter"~\\%
\lstinline[style=tlsf_keyword]~  DESCRIPTION~%
\lstinline[style=tlsf_default]~: ~%
\lstinline[style=tlsf_istring]~"Component: Encode"~\\%
\lstinline[style=tlsf_keyword]~  SEMANTICS~%
\lstinline[style=tlsf_default]~:   ~%
\lstinline[style=tlsf_semanti]~Mealy~\\%
\lstinline[style=tlsf_keyword]~  TARGET~%
\lstinline[style=tlsf_default]~:      ~%
\lstinline[style=tlsf_semanti]~Mealy~\\%
\lstinline[style=tlsf_default]~}~\\%
\lstinline[style=tlsf_keyword]~GLOBAL ~%
\lstinline[style=tlsf_default]~{~\\%
\lstinline[style=tlsf_keyword]~  PARAMETERS ~%
\lstinline[style=tlsf_default]~{~\\%
\lstinline[style=tlsf_variabl]~    n~%
\lstinline[style=tlsf_default]~ = 2;~\\%
\lstinline[style=tlsf_default]~  }~\\%
\lstinline[style=tlsf_keyword]~  DEFINITIONS~%
\lstinline[style=tlsf_default]~ {~\\%
\lstinline[style=tlsf_comment]~    // mutual exclusion~\\%
\lstinline[style=tlsf_variabl]~    mutual~%
\lstinline[style=tlsf_default]~(b) =~\\%
\lstinline[style=tlsf_operato]~      ||~%
\lstinline[style=tlsf_default]~[i ~%
\lstinline[style=tlsf_operato]~IN~%
\lstinline[style=tlsf_default]~ {0, 1 .. (~%
\lstinline[style=tlsf_operato]~SIZEOF ~%
\lstinline[style=tlsf_default]~b)~%
\lstinline[style=tlsf_operato]~ - ~%
\lstinline[style=tlsf_default]~1}]~\\%
\lstinline[style=tlsf_operato]~        &&~%
\lstinline[style=tlsf_default]~[j ~%
\lstinline[style=tlsf_operato]~IN~%
\lstinline[style=tlsf_default]~ {0, 1 .. (~%
\lstinline[style=tlsf_operato]~SIZEOF ~%
\lstinline[style=tlsf_default]~b)~%
\lstinline[style=tlsf_operato]~ - ~%
\lstinline[style=tlsf_default]~1} ~%
\lstinline[style=tlsf_operato]~(\)~%
\lstinline[style=tlsf_default]~ {i}]~\\%
\lstinline[style=tlsf_operato]~          !~%
\lstinline[style=tlsf_default]~(b[i] ~%
\lstinline[style=tlsf_operato]~&&~%
\lstinline[style=tlsf_default]~ b[j]);~\\%
\lstinline[style=tlsf_comment]~    // checks whether a bus encodes the numerical~\\%
\lstinline[style=tlsf_comment]~    // value v in binary~\\%
\lstinline[style=tlsf_variabl]~    value~%
\lstinline[style=tlsf_default]~(bus,v) = value'(bus,v,0,~
\lstinline[style=tlsf_operato]~SIZEOF~
\lstinline[style=tlsf_default]~ bus);~\\%
\lstinline[style=tlsf_variabl]~    value'~%
\lstinline[style=tlsf_default]~(bus,v,i,j) =~\\%
\lstinline[style=tlsf_default]~      i ~%
\lstinline[style=tlsf_operato]~>=~%
\lstinline[style=tlsf_default]~ j        : ~%
\lstinline[style=tlsf_keyword]~true~\\%
\lstinline[style=tlsf_default]~      bit(v,i) ~%
\lstinline[style=tlsf_operato]~==~%
\lstinline[style=tlsf_default]~ 1 : value'(bus,v,i~%
\lstinline[style=tlsf_operato]~+~%
\lstinline[style=tlsf_default]~1,j)~\\%
\lstinline[style=tlsf_operato]~                      &&~%
\lstinline[style=tlsf_default]~ bus[i]~\\%
\lstinline[style=tlsf_keyword]~      otherwise~%
\lstinline[style=tlsf_default]~     : value'(bus,v,i~%
\lstinline[style=tlsf_operato]~+~%
\lstinline[style=tlsf_default]~1,j)~\\%
\lstinline[style=tlsf_operato]~                      && !~%
\lstinline[style=tlsf_default]~bus[i];~\\%
\lstinline[style=tlsf_comment]~    // returns the i-th bit of the numerical~\\%
\lstinline[style=tlsf_comment]~    // value v~\\%
\lstinline[style=tlsf_variabl]~    bit~%
\lstinline[style=tlsf_default]~(v,i) =~\\%
\lstinline[style=tlsf_default]~      i ~%
\lstinline[style=tlsf_operato]~<=~%
\lstinline[style=tlsf_default]~ 0    : v ~%
\lstinline[style=tlsf_operato]~
\lstinline[style=tlsf_default]~ 2~\\%
\lstinline[style=tlsf_keyword]~      otherwise~%
\lstinline[style=tlsf_default]~ : bit(v~%
\lstinline[style=tlsf_operato]~/~%
\lstinline[style=tlsf_default]~2,i~%
\lstinline[style=tlsf_operato]~-~%
\lstinline[style=tlsf_default]~1);~\\%
\lstinline[style=tlsf_comment]~    // discrete logarithm~\\%
\lstinline[style=tlsf_variabl]~    log2~%
\lstinline[style=tlsf_default]~(x) =~\\%
\lstinline[style=tlsf_default]~      x ~%
\lstinline[style=tlsf_operato]~<=~%
\lstinline[style=tlsf_default]~ 1    : 1~\\%
\lstinline[style=tlsf_keyword]~      otherwise~%
\lstinline[style=tlsf_default]~ : 1 ~%
\lstinline[style=tlsf_operato]~+~%
\lstinline[style=tlsf_default]~ log2(x~%
\lstinline[style=tlsf_operato]~/~%
\lstinline[style=tlsf_default]~2);~\\%
\lstinline[style=tlsf_default]~  }~\\%
\lstinline[style=tlsf_default]~}~\\%
\lstinline[style=tlsf_keyword]~MAIN~%
\lstinline[style=tlsf_default]~ {~\\%
\lstinline[style=tlsf_keyword]~  INPUTS~%
\lstinline[style=tlsf_default]~ {~\\%
\lstinline[style=tlsf_variabl]~    HREADY~%
\lstinline[style=tlsf_default]~;~\\%
\lstinline[style=tlsf_variabl]~    HGRANT~%
\lstinline[style=tlsf_default]~[n];~\\%
\lstinline[style=tlsf_default]~  }~\\%
\lstinline[style=tlsf_keyword]~  OUTPUTS~%
\lstinline[style=tlsf_default]~ {~\\%
\lstinline[style=tlsf_comment]~    // the output is encoded in binary~\\%
\lstinline[style=tlsf_variabl]~    HMASTER~%
\lstinline[style=tlsf_default]~[log2(n~%
\lstinline[style=tlsf_operato]~-~%
\lstinline[style=tlsf_default]~1)];~\\%
\lstinline[style=tlsf_default]~  }~\\%
\lstinline[style=tlsf_keyword]~  REQUIRE~%
\lstinline[style=tlsf_default]~ {~\\%
\lstinline[style=tlsf_comment]~    // a every time exactely one grant is high~\\%
\lstinline[style=tlsf_default]~    mutual(HGRANT) ~%
\lstinline[style=tlsf_operato]~&& ||~%
\lstinline[style=tlsf_default]~[0 ~%
\lstinline[style=tlsf_operato]~<=~%
\lstinline[style=tlsf_default]~ i ~%
\lstinline[style=tlsf_operato]~<~%
\lstinline[style=tlsf_default]~ n] HGRANT[i];~\\%
\lstinline[style=tlsf_default]~  }~\\%
\lstinline[style=tlsf_keyword]~  ASSERT~%
\lstinline[style=tlsf_default]~ {~\\%
\lstinline[style=tlsf_comment]~    // output the binary encoding of i, whenever~\\%
\lstinline[style=tlsf_comment]~    // i is granted and HREADY is high~\\%
\lstinline[style=tlsf_operato]~    &&~%
\lstinline[style=tlsf_default]~[0 ~%
\lstinline[style=tlsf_operato]~<=~%
\lstinline[style=tlsf_default]~ i ~%
\lstinline[style=tlsf_operato]~<~%
\lstinline[style=tlsf_default]~ n] (HREADY ~%
\lstinline[style=tlsf_operato]~->~\\%
\lstinline[style=tlsf_default]~      (~%
\lstinline[style=tlsf_operato]~X~%
\lstinline[style=tlsf_default]~ value(HMASTER,i) ~%
\lstinline[style=tlsf_operato]~<->~%
\lstinline[style=tlsf_default]~ HGRANT[i]));~\\%
\lstinline[style=tlsf_comment]~    // when HREADY is low, the value is copied~\\%
\lstinline[style=tlsf_operato]~    !~%
\lstinline[style=tlsf_default]~HREADY ~%
\lstinline[style=tlsf_operato]~->~%
\lstinline[style=tlsf_operato]~ &&~%
\lstinline[style=tlsf_default]~[0 ~%
\lstinline[style=tlsf_operato]~<=~%
\lstinline[style=tlsf_default]~ i ~%
\lstinline[style=tlsf_operato]~<~%
\lstinline[style=tlsf_default]~ log2(n~%
\lstinline[style=tlsf_operato]~-~%
\lstinline[style=tlsf_default]~1)]~\\%
\lstinline[style=tlsf_default]~      (~%
\lstinline[style=tlsf_operato]~X~%
\lstinline[style=tlsf_default]~ HMASTER[i] ~%
\lstinline[style=tlsf_operato]~<->~%
\lstinline[style=tlsf_default]~ HMASTER[i]);~\\%
\lstinline[style=tlsf_default]~  }~\\%
\lstinline[style=tlsf_default]~}~
\end{minipage}
\end{lrbox}

\newsavebox{\cShift}
\begin{lrbox}{\cShift}
\begin{minipage}{0.9\textwidth}
\scriptsize
\lstinline[style=tlsf_keyword]~INFO~%
\lstinline[style=tlsf_default]~ {~\\%
\lstinline[style=tlsf_keyword]~  TITLE~%
\lstinline[style=tlsf_default]~:       ~%
\lstinline[style=tlsf_istring]~"AMBA AHB Arbiter"~\\%
\lstinline[style=tlsf_keyword]~  DESCRIPTION~%
\lstinline[style=tlsf_default]~: ~%
\lstinline[style=tlsf_istring]~"Component: Shift"~\\%
\lstinline[style=tlsf_keyword]~  SEMANTICS~%
\lstinline[style=tlsf_default]~:   ~%
\lstinline[style=tlsf_semanti]~Mealy~\\%
\lstinline[style=tlsf_keyword]~  TARGET~%
\lstinline[style=tlsf_default]~:      ~%
\lstinline[style=tlsf_semanti]~Mealy~\\%
\lstinline[style=tlsf_default]~}~\\%
\lstinline[style=tlsf_keyword]~MAIN~%
\lstinline[style=tlsf_default]~ {~\\%
\lstinline[style=tlsf_keyword]~  INPUTS~%
\lstinline[style=tlsf_default]~ { ~%
\lstinline[style=tlsf_variabl]~HREADY~%
\lstinline[style=tlsf_default]~; ~%
\lstinline[style=tlsf_variabl]~LOCKED~%
\lstinline[style=tlsf_default]~;~%
\lstinline[style=tlsf_default]~ }~\\%
\lstinline[style=tlsf_keyword]~  OUTPUTS~%
\lstinline[style=tlsf_default]~ { ~%
\lstinline[style=tlsf_variabl]~HMASTLOCK~%
\lstinline[style=tlsf_default]~;~%
\lstinline[style=tlsf_default]~ }~\\%
\lstinline[style=tlsf_keyword]~  ASSERT~%
\lstinline[style=tlsf_default]~ {~\\%
\lstinline[style=tlsf_comment]~    // if HREADY is high, the component copies LOCKED to HMASTLOCK, shifted by one time step~\\%
\lstinline[style=tlsf_default]~    HREADY ~%
\lstinline[style=tlsf_operato]~->~%
\lstinline[style=tlsf_default]~ (~%
\lstinline[style=tlsf_operato]~X~%
\lstinline[style=tlsf_default]~ HMASTLOCK ~%
\lstinline[style=tlsf_operato]~<->~%
\lstinline[style=tlsf_default]~ LOCKED);~\\%
\lstinline[style=tlsf_comment]~    // if HREADY is low, the old value of HMASTLOCK is copied~\\%
\lstinline[style=tlsf_operato]~    !~%
\lstinline[style=tlsf_default]~HREADY ~%
\lstinline[style=tlsf_operato]~->~%
\lstinline[style=tlsf_default]~ (~%
\lstinline[style=tlsf_operato]~X~%
\lstinline[style=tlsf_default]~ HMASTLOCK ~%
\lstinline[style=tlsf_operato]~<->~%
\lstinline[style=tlsf_default]~ HMASTLOCK);~\\%
\lstinline[style=tlsf_default]~  }~\\%
\lstinline[style=tlsf_default]~}~
\end{minipage}
\end{lrbox}

\newsavebox{\cTincr}
\begin{lrbox}{\cTincr}
\begin{minipage}{0.9\textwidth}
\scriptsize
\lstinline[style=tlsf_keyword]~INFO~%
\lstinline[style=tlsf_default]~ {~\\%
\lstinline[style=tlsf_keyword]~  TITLE~%
\lstinline[style=tlsf_default]~:       ~%
\lstinline[style=tlsf_istring]~"AMBA AHB Arbiter"~\\%
\lstinline[style=tlsf_keyword]~  DESCRIPTION~%
\lstinline[style=tlsf_default]~: ~%
\lstinline[style=tlsf_istring]~"Component: TIncr"~\\%
\lstinline[style=tlsf_keyword]~  SEMANTICS~%
\lstinline[style=tlsf_default]~:   ~%
\lstinline[style=tlsf_semanti]~Mealy~\\%
\lstinline[style=tlsf_keyword]~  TARGET~%
\lstinline[style=tlsf_default]~:      ~%
\lstinline[style=tlsf_semanti]~Mealy~\\%
\lstinline[style=tlsf_default]~}~\\%
\lstinline[style=tlsf_keyword]~MAIN~%
\lstinline[style=tlsf_default]~ {~\\%
\lstinline[style=tlsf_keyword]~  INPUTS~%
\lstinline[style=tlsf_default]~ { ~%
\lstinline[style=tlsf_variabl]~INCR~%
\lstinline[style=tlsf_default]~; ~%
\lstinline[style=tlsf_variabl]~HREADY~%
\lstinline[style=tlsf_default]~; ~%
\lstinline[style=tlsf_variabl]~LOCKED~%
\lstinline[style=tlsf_default]~; ~%
\lstinline[style=tlsf_variabl]~DECIDE~%
\lstinline[style=tlsf_default]~; ~%
\lstinline[style=tlsf_variabl]~BUSREQ~%
\lstinline[style=tlsf_default]~; ~%
\lstinline[style=tlsf_default]~}~\\%
\lstinline[style=tlsf_keyword]~  OUTPUTS~%
\lstinline[style=tlsf_default]~ { ~%
\lstinline[style=tlsf_variabl]~READY1~%
\lstinline[style=tlsf_default]~;~%
\lstinline[style=tlsf_default]~ }~\\%
\lstinline[style=tlsf_keyword]~  INITIALLY~%
\lstinline[style=tlsf_default]~ { ~%
\lstinline[style=tlsf_operato]~!~%
\lstinline[style=tlsf_default]~DECIDE;~%
\lstinline[style=tlsf_default]~ }~\\%
\lstinline[style=tlsf_keyword]~  PRESET~%
\lstinline[style=tlsf_default]~ { ~%
\lstinline[style=tlsf_default]~READY1;~%
\lstinline[style=tlsf_default]~ }~\\%
\lstinline[style=tlsf_keyword]~  REQUIRE~%
\lstinline[style=tlsf_default]~ {~\\%
\lstinline[style=tlsf_comment]~    // decisions are only taken if the component is ready~\\%
\lstinline[style=tlsf_operato]~    !~%
\lstinline[style=tlsf_default]~READY1 ~%
\lstinline[style=tlsf_operato]~-> X !~%
\lstinline[style=tlsf_default]~DECIDE;~\\%
\lstinline[style=tlsf_default]~  }~\\%
\lstinline[style=tlsf_keyword]~  ASSUME~%
\lstinline[style=tlsf_default]~ {~\\%
\lstinline[style=tlsf_comment]~    // slaves and masters cannot block the bus~\\%
\lstinline[style=tlsf_operato]~    G F ~%
\lstinline[style=tlsf_default]~HREADY ~%
\lstinline[style=tlsf_operato]~&& G F !~%
\lstinline[style=tlsf_default]~BUSREQ;~\\%
\lstinline[style=tlsf_default]~  }~\\%
\lstinline[style=tlsf_keyword]~  ASSERT~%
\lstinline[style=tlsf_default]~ {~\\%
\lstinline[style=tlsf_comment]~    // for each incremental, locked transmission, the bus is locked as long as requested~\\%
\lstinline[style=tlsf_default]~    DECIDE ~%
\lstinline[style=tlsf_operato]~->~\\%
\lstinline[style=tlsf_operato]~      X~%
\lstinline[style=tlsf_default]~[2] (((INCR ~%
\lstinline[style=tlsf_operato]~&&~%
\lstinline[style=tlsf_default]~ LOCKED) ~%
\lstinline[style=tlsf_operato]~->~%
\lstinline[style=tlsf_default]~ (~%
\lstinline[style=tlsf_operato]~!~%
\lstinline[style=tlsf_default]~READY1 ~%
\lstinline[style=tlsf_operato]~W~%
\lstinline[style=tlsf_default]~ (HREADY ~%
\lstinline[style=tlsf_operato]~&& !~
\lstinline[style=tlsf_default]~BUSREQ))) ~%
\lstinline[style=tlsf_operato]~&&~\\%
\lstinline[style=tlsf_default]~            (~%
\lstinline[style=tlsf_operato]~!~%
\lstinline[style=tlsf_default]~(INCR ~%
\lstinline[style=tlsf_operato]~&&~%
\lstinline[style=tlsf_default]~ LOCKED) ~%
\lstinline[style=tlsf_operato]~->~%
\lstinline[style=tlsf_default]~ READY1));~\\%
\lstinline[style=tlsf_comment]~    // the component stays ready as long as there is no decision~\\%
\lstinline[style=tlsf_default]~    READY ~%
\lstinline[style=tlsf_operato]~&& X !~%
\lstinline[style=tlsf_default]~DECIDE ~%
\lstinline[style=tlsf_operato]~-> X~%
\lstinline[style=tlsf_default]~ READY1;~\\%
\lstinline[style=tlsf_comment]~    // if there is a decision the component blocks the bus for at least two time steps~\\%
\lstinline[style=tlsf_default]~    READY1 ~%
\lstinline[style=tlsf_operato]~&& X~%
\lstinline[style=tlsf_default]~ DECIDE ~%
\lstinline[style=tlsf_operato]~-> G~%
\lstinline[style=tlsf_default]~[1:2] ~%
\lstinline[style=tlsf_operato]~!~%
\lstinline[style=tlsf_default]~ READY1;~\\%
\lstinline[style=tlsf_default]~  }~\\%
\lstinline[style=tlsf_default]~}~
\end{minipage}
\end{lrbox}

\newsavebox{\cSingle}
\begin{lrbox}{\cSingle}
\begin{minipage}{0.9\textwidth}
\scriptsize
\lstinline[style=tlsf_keyword]~INFO~%
\lstinline[style=tlsf_default]~ {~\\%
\lstinline[style=tlsf_keyword]~  TITLE~%
\lstinline[style=tlsf_default]~:       ~%
\lstinline[style=tlsf_istring]~"AMBA AHB Arbiter"~\\%
\lstinline[style=tlsf_keyword]~  DESCRIPTION~%
\lstinline[style=tlsf_default]~: ~%
\lstinline[style=tlsf_istring]~"Component: TSingle"~\\%
\lstinline[style=tlsf_keyword]~  SEMANTICS~%
\lstinline[style=tlsf_default]~:   ~%
\lstinline[style=tlsf_semanti]~Mealy~\\%
\lstinline[style=tlsf_keyword]~  TARGET~%
\lstinline[style=tlsf_default]~:      ~%
\lstinline[style=tlsf_semanti]~Mealy~\\%
\lstinline[style=tlsf_default]~}~\\%
\lstinline[style=tlsf_keyword]~MAIN~%
\lstinline[style=tlsf_default]~ {~\\%
\lstinline[style=tlsf_keyword]~  INPUTS~%
\lstinline[style=tlsf_default]~ { ~%
\lstinline[style=tlsf_variabl]~SINGLE~%
\lstinline[style=tlsf_default]~; ~%
\lstinline[style=tlsf_variabl]~HREADY~%
\lstinline[style=tlsf_default]~; ~%
\lstinline[style=tlsf_variabl]~LOCKED~%
\lstinline[style=tlsf_default]~; ~%
\lstinline[style=tlsf_variabl]~DECIDE~%
\lstinline[style=tlsf_default]~; ~%
\lstinline[style=tlsf_default]~}~\\%
\lstinline[style=tlsf_keyword]~  OUTPUTS~%
\lstinline[style=tlsf_default]~ { ~%
\lstinline[style=tlsf_variabl]~READY3~%
\lstinline[style=tlsf_default]~;~%
\lstinline[style=tlsf_default]~ }~\\%
\lstinline[style=tlsf_keyword]~  INITIALLY~%
\lstinline[style=tlsf_default]~ {~\\%
\lstinline[style=tlsf_comment]~    // initially no decision is taken~\\%
\lstinline[style=tlsf_operato]~    !~%
\lstinline[style=tlsf_default]~DECIDE;~\\%
\lstinline[style=tlsf_default]~  }~\\%
\lstinline[style=tlsf_keyword]~  PRESET~%
\lstinline[style=tlsf_default]~ {~\\%
\lstinline[style=tlsf_comment]~    // at startup, the component is ready~\\%
\lstinline[style=tlsf_default]~    READY3;~\\%
\lstinline[style=tlsf_default]~  }~\\%
\lstinline[style=tlsf_keyword]~  REQUIRE~%
\lstinline[style=tlsf_default]~ {~\\%
\lstinline[style=tlsf_comment]~    // decisions are only taken if the component is ready~\\%
\lstinline[style=tlsf_operato]~    !~%
\lstinline[style=tlsf_default]~READY3 ~%
\lstinline[style=tlsf_operato]~-> X !~%
\lstinline[style=tlsf_default]~DECIDE;~\\%
\lstinline[style=tlsf_default]~  }~\\%
\lstinline[style=tlsf_keyword]~  ASSUME~%
\lstinline[style=tlsf_default]~ {~\\%
\lstinline[style=tlsf_comment]~    // a slave cannot block the bus~\\%
\lstinline[style=tlsf_operato]~    G F ~%
\lstinline[style=tlsf_default]~HREADY ~\\%
\lstinline[style=tlsf_default]~  }~\\%
\lstinline[style=tlsf_keyword]~  ASSERT~%
\lstinline[style=tlsf_default]~ {~\\%
\lstinline[style=tlsf_comment]~    // for each single, locked transmission, the bus is locked for one time step~\\%
\lstinline[style=tlsf_default]~    DECIDE ~%
\lstinline[style=tlsf_operato]~->~\\%
\lstinline[style=tlsf_operato]~      X~%
\lstinline[style=tlsf_default]~[2] (((SINGLE ~%
\lstinline[style=tlsf_operato]~&&~%
\lstinline[style=tlsf_default]~ LOCKED) ~%
\lstinline[style=tlsf_operato]~->~%
\lstinline[style=tlsf_default]~ (~%
\lstinline[style=tlsf_operato]~!~%
\lstinline[style=tlsf_default]~READY3 ~%
\lstinline[style=tlsf_operato]~U~%
\lstinline[style=tlsf_default]~ (HREADY ~%
\lstinline[style=tlsf_operato]~&& !~
\lstinline[style=tlsf_default]~READY3 ~%
\lstinline[style=tlsf_operato]~&& X~%
\lstinline[style=tlsf_default]~ READY3))) ~%
\lstinline[style=tlsf_operato]~&&~\\%
\lstinline[style=tlsf_default]~            (~%
\lstinline[style=tlsf_operato]~!~%
\lstinline[style=tlsf_default]~(SINGLE ~%
\lstinline[style=tlsf_operato]~&&~%
\lstinline[style=tlsf_default]~ LOCKED) ~%
\lstinline[style=tlsf_operato]~->~%
\lstinline[style=tlsf_default]~ READY3));~\\%
\lstinline[style=tlsf_comment]~    // the component stays ready as long as there is no decision~\\%
\lstinline[style=tlsf_default]~    READY3 ~%
\lstinline[style=tlsf_operato]~&& X !~%
\lstinline[style=tlsf_default]~DECIDE ~%
\lstinline[style=tlsf_operato]~-> X~%
\lstinline[style=tlsf_default]~ READY3;~\\%
\lstinline[style=tlsf_comment]~    // if there is a decision the component blocks the bus for at least two time steps~\\%
\lstinline[style=tlsf_default]~    READY3 ~%
\lstinline[style=tlsf_operato]~&& X~%
\lstinline[style=tlsf_default]~ DECIDE ~%
\lstinline[style=tlsf_operato]~-> G~%
\lstinline[style=tlsf_default]~[1:2] ~%
\lstinline[style=tlsf_operato]~!~%
\lstinline[style=tlsf_default]~ READY3;~\\%
\lstinline[style=tlsf_default]~  }~\\%
\lstinline[style=tlsf_default]~}~
\end{minipage}
\end{lrbox}

\newsavebox{\cBurst}
\begin{lrbox}{\cBurst}
\begin{minipage}{0.9\textwidth}
\scriptsize
\lstinline[style=tlsf_keyword]~INFO~%
\lstinline[style=tlsf_default]~ {~\\%
\lstinline[style=tlsf_keyword]~  TITLE~%
\lstinline[style=tlsf_default]~:       ~%
\lstinline[style=tlsf_istring]~"AMBA AHB Arbiter"~\\%
\lstinline[style=tlsf_keyword]~  DESCRIPTION~%
\lstinline[style=tlsf_default]~: ~%
\lstinline[style=tlsf_istring]~"Component: TBurst4"~\\%
\lstinline[style=tlsf_keyword]~  SEMANTICS~%
\lstinline[style=tlsf_default]~:   ~%
\lstinline[style=tlsf_semanti]~Mealy~\\%
\lstinline[style=tlsf_keyword]~  TARGET~%
\lstinline[style=tlsf_default]~:      ~%
\lstinline[style=tlsf_semanti]~Mealy~\\%
\lstinline[style=tlsf_default]~}~\\%
\lstinline[style=tlsf_keyword]~MAIN~%
\lstinline[style=tlsf_default]~ {~\\%
\lstinline[style=tlsf_keyword]~  INPUTS~%
\lstinline[style=tlsf_default]~ { ~%
\lstinline[style=tlsf_variabl]~BURST4~%
\lstinline[style=tlsf_default]~; ~%
\lstinline[style=tlsf_variabl]~HREADY~%
\lstinline[style=tlsf_default]~; ~%
\lstinline[style=tlsf_variabl]~LOCKED~%
\lstinline[style=tlsf_default]~; ~%
\lstinline[style=tlsf_variabl]~DECIDE~%
\lstinline[style=tlsf_default]~; ~%
\lstinline[style=tlsf_default]~}~\\%
\lstinline[style=tlsf_keyword]~  OUTPUTS~%
\lstinline[style=tlsf_default]~ { ~%
\lstinline[style=tlsf_variabl]~READY2~%
\lstinline[style=tlsf_default]~;~%
\lstinline[style=tlsf_default]~ }~\\%
\lstinline[style=tlsf_keyword]~  INITIALLY~%
\lstinline[style=tlsf_default]~ { ~%
\lstinline[style=tlsf_operato]~!~%
\lstinline[style=tlsf_default]~DECIDE;~%
\lstinline[style=tlsf_default]~ }~\\%
\lstinline[style=tlsf_keyword]~  PRESET~%
\lstinline[style=tlsf_default]~ { ~%
\lstinline[style=tlsf_default]~READY2;~%
\lstinline[style=tlsf_default]~ }~\\%
\lstinline[style=tlsf_keyword]~  REQUIRE~%
\lstinline[style=tlsf_default]~ {~\\%
\lstinline[style=tlsf_comment]~    // decisions are only taken if the component is ready~\\%
\lstinline[style=tlsf_operato]~    !~%
\lstinline[style=tlsf_default]~READY2 ~%
\lstinline[style=tlsf_operato]~-> X !~%
\lstinline[style=tlsf_default]~DECIDE;~\\%
\lstinline[style=tlsf_default]~  }~\\%
\lstinline[style=tlsf_keyword]~  ASSUME~%
\lstinline[style=tlsf_default]~ {~\\%
\lstinline[style=tlsf_comment]~    // a slave block the bus~\\%
\lstinline[style=tlsf_operato]~    G F ~%
\lstinline[style=tlsf_default]~HREADY;~\\%
\lstinline[style=tlsf_default]~  }~\\%
\lstinline[style=tlsf_keyword]~  ASSERT~%
\lstinline[style=tlsf_default]~ {~\\%
\lstinline[style=tlsf_comment]~    // for each burst4, locked transmission, the bus is locked for four time steps~\\%
\lstinline[style=tlsf_default]~    DECIDE ~%
\lstinline[style=tlsf_operato]~->~\\%
\lstinline[style=tlsf_operato]~      X~%
\lstinline[style=tlsf_default]~[2] (((BURST4 ~%
\lstinline[style=tlsf_operato]~&&~%
\lstinline[style=tlsf_default]~ LOCKED) ~%
\lstinline[style=tlsf_operato]~->~%
\lstinline[style=tlsf_default]~ (~%
\lstinline[style=tlsf_operato]~!~%
\lstinline[style=tlsf_default]~READY2 ~%
\lstinline[style=tlsf_operato]~U~%
\lstinline[style=tlsf_default]~ (HREADY ~%
\lstinline[style=tlsf_operato]~&& !~
\lstinline[style=tlsf_default]~READY2 ~%
\lstinline[style=tlsf_operato]~&& X~%
\lstinline[style=tlsf_default]~ (~%
\lstinline[style=tlsf_operato]~!~%
\lstinline[style=tlsf_default]~READY2 ~%
\lstinline[style=tlsf_operato]~U~%
\lstinline[style=tlsf_default]~ (HREADY ~%
\lstinline[style=tlsf_operato]~&&~\\%
\lstinline[style=tlsf_operato]~             !~%
\lstinline[style=tlsf_default]~READY2 ~%
\lstinline[style=tlsf_operato]~&& X~%
\lstinline[style=tlsf_default]~ (~%
\lstinline[style=tlsf_operato]~!~%
\lstinline[style=tlsf_default]~READY2 ~%
\lstinline[style=tlsf_operato]~U~%
\lstinline[style=tlsf_default]~ (HREADY ~%
\lstinline[style=tlsf_operato]~&& !~%
\lstinline[style=tlsf_default]~READY2 ~%
\lstinline[style=tlsf_operato]~&& X~%
\lstinline[style=tlsf_default]~ (~%
\lstinline[style=tlsf_operato]~!~%
\lstinline[style=tlsf_default]~READY2 ~%
\lstinline[style=tlsf_operato]~U~%
\lstinline[style=tlsf_default]~ (HREADY ~%
\lstinline[style=tlsf_operato]~&&~\\%
\lstinline[style=tlsf_operato]~             !~%
\lstinline[style=tlsf_default]~READY2 ~%
\lstinline[style=tlsf_operato]~&& X~%
\lstinline[style=tlsf_default]~READY2))))))))) ~%
\lstinline[style=tlsf_operato]~&&~%
\lstinline[style=tlsf_default]~ (~%
\lstinline[style=tlsf_operato]~!~%
\lstinline[style=tlsf_default]~(BURST4 ~%
\lstinline[style=tlsf_operato]~&&~%
\lstinline[style=tlsf_default]~ LOCKED) ~%
\lstinline[style=tlsf_operato]~->~%
\lstinline[style=tlsf_default]~ READY2)) ~\\%
\lstinline[style=tlsf_comment]~    // the component stays ready as long as there is no decision~\\%
\lstinline[style=tlsf_default]~    READY2 ~%
\lstinline[style=tlsf_operato]~&& X !~%
\lstinline[style=tlsf_default]~DECIDE ~%
\lstinline[style=tlsf_operato]~-> X~%
\lstinline[style=tlsf_default]~ READY2;~\\%
\lstinline[style=tlsf_comment]~    // if there is a decision the component blocks the bus for at least two time steps~\\%
\lstinline[style=tlsf_default]~    READY2 ~%
\lstinline[style=tlsf_operato]~&& X~%
\lstinline[style=tlsf_default]~ DECIDE ~%
\lstinline[style=tlsf_operato]~-> G~%
\lstinline[style=tlsf_default]~[1:2] ~%
\lstinline[style=tlsf_operato]~!~%
\lstinline[style=tlsf_default]~ READY2;~\\%
\lstinline[style=tlsf_default]~  }~\\%
\lstinline[style=tlsf_default]~}~
\end{minipage}
\end{lrbox}

\newsavebox{\cLock}
\begin{lrbox}{\cLock}
\begin{minipage}{0.9\textwidth}
\scriptsize
\lstinline[style=tlsf_keyword]~INFO~%
\lstinline[style=tlsf_default]~ {~\\%
\lstinline[style=tlsf_keyword]~  TITLE~%
\lstinline[style=tlsf_default]~:       ~%
\lstinline[style=tlsf_istring]~"AMBA AHB Arbiter"~\\%
\lstinline[style=tlsf_keyword]~  DESCRIPTION~%
\lstinline[style=tlsf_default]~: ~%
\lstinline[style=tlsf_istring]~"Component: Lock"~\\%
\lstinline[style=tlsf_keyword]~  SEMANTICS~%
\lstinline[style=tlsf_default]~:   ~%
\lstinline[style=tlsf_semanti]~Mealy~\\%
\lstinline[style=tlsf_keyword]~  TARGET~%
\lstinline[style=tlsf_default]~:      ~%
\lstinline[style=tlsf_semanti]~Mealy~\\%
\lstinline[style=tlsf_default]~}~\\%
\lstinline[style=tlsf_keyword]~GLOBAL ~%
\lstinline[style=tlsf_default]~{~\\%
\lstinline[style=tlsf_keyword]~  PARAMETERS ~%
\lstinline[style=tlsf_default]~{~\\%
\lstinline[style=tlsf_variabl]~    n~%
\lstinline[style=tlsf_default]~ = 2;~\\%
\lstinline[style=tlsf_default]~  }~\\%
\lstinline[style=tlsf_keyword]~  DEFINITIONS~%
\lstinline[style=tlsf_default]~ {~\\%
\lstinline[style=tlsf_comment]~    // mutual exclusion~\\%
\lstinline[style=tlsf_variabl]~    mutual~%
\lstinline[style=tlsf_default]~(b) =~\\%
\lstinline[style=tlsf_operato]~      ||~%
\lstinline[style=tlsf_default]~[i ~%
\lstinline[style=tlsf_operato]~IN~%
\lstinline[style=tlsf_default]~ {0, 1 .. (~%
\lstinline[style=tlsf_operato]~SIZEOF ~%
\lstinline[style=tlsf_default]~b)~%
\lstinline[style=tlsf_operato]~ - ~%
\lstinline[style=tlsf_default]~1}]~\\%
\lstinline[style=tlsf_operato]~        &&~%
\lstinline[style=tlsf_default]~[j ~%
\lstinline[style=tlsf_operato]~IN~%
\lstinline[style=tlsf_default]~ {0, 1 .. (~%
\lstinline[style=tlsf_operato]~SIZEOF ~%
\lstinline[style=tlsf_default]~b)~%
\lstinline[style=tlsf_operato]~ - ~%
\lstinline[style=tlsf_default]~1} ~%
\lstinline[style=tlsf_operato]~(\)~%
\lstinline[style=tlsf_default]~ {i}]~\\%
\lstinline[style=tlsf_operato]~          !~%
\lstinline[style=tlsf_default]~(b[i] ~%
\lstinline[style=tlsf_operato]~&&~%
\lstinline[style=tlsf_default]~ b[j]);~\\%
\lstinline[style=tlsf_default]~  }~\\%
\lstinline[style=tlsf_default]~}~\\%
\lstinline[style=tlsf_keyword]~MAIN~%
\lstinline[style=tlsf_default]~ {~\\%
\lstinline[style=tlsf_keyword]~  INPUTS~%
\lstinline[style=tlsf_default]~ {~\\%
\lstinline[style=tlsf_variabl]~    DECIDE~%
\lstinline[style=tlsf_default]~;~\\%
\lstinline[style=tlsf_variabl]~    HGRANT~%
\lstinline[style=tlsf_default]~[n];~\\%
\lstinline[style=tlsf_variabl]~    HLOCK~%
\lstinline[style=tlsf_default]~[n];~\\%
\lstinline[style=tlsf_default]~  }~\\%
\lstinline[style=tlsf_keyword]~  OUTPUTS~%
\lstinline[style=tlsf_default]~ {~\\%
\lstinline[style=tlsf_variabl]~    LOCKED~%
\lstinline[style=tlsf_default]~;~\\%
\lstinline[style=tlsf_default]~  }~\\%
\lstinline[style=tlsf_keyword]~  REQUIRE~%
\lstinline[style=tlsf_default]~ {~\\%
\lstinline[style=tlsf_comment]~    // a every time exactely one grant is high~\\%
\lstinline[style=tlsf_default]~    mutual(HGRANT) ~%
\lstinline[style=tlsf_operato]~&& ||~%
\lstinline[style=tlsf_default]~[0 ~%
\lstinline[style=tlsf_operato]~<=~%
\lstinline[style=tlsf_default]~ i ~%
\lstinline[style=tlsf_operato]~<~%
\lstinline[style=tlsf_default]~ n] HGRANT[i];~\\%
\lstinline[style=tlsf_default]~  }~\\%
\lstinline[style=tlsf_keyword]~  ASSERT~%
\lstinline[style=tlsf_default]~ {~\\%
\lstinline[style=tlsf_comment]~    // whenever a decicion is taken, the LOCKED signal is updated to~\\%
\lstinline[style=tlsf_comment]~    // the HLOCK value of the granted master~\\%
\lstinline[style=tlsf_operato]~    &&~%
\lstinline[style=tlsf_default]~[0 ~%
\lstinline[style=tlsf_operato]~<=~%
\lstinline[style=tlsf_default]~ i ~%
\lstinline[style=tlsf_operato]~<~%
\lstinline[style=tlsf_default]~ n] (DECIDE ~%
\lstinline[style=tlsf_operato]~&& X~%
\lstinline[style=tlsf_default]~ HGRANT[i] ~%
\lstinline[style=tlsf_operato]~->~%
\lstinline[style=tlsf_default]~ (~%
\lstinline[style=tlsf_operato]~X~%
\lstinline[style=tlsf_default]~ LOCKED ~%
\lstinline[style=tlsf_operato]~<-> X~%
\lstinline[style=tlsf_default]~ HLOCK[i]));~\\%
\lstinline[style=tlsf_comment]~    // otherwise, the value is copied~\\%
\lstinline[style=tlsf_operato]~    !~%
\lstinline[style=tlsf_default]~DECIDE ~%
\lstinline[style=tlsf_operato]~->~%
\lstinline[style=tlsf_default]~ (~%
\lstinline[style=tlsf_operato]~X~%
\lstinline[style=tlsf_default]~ LOCKED ~%
\lstinline[style=tlsf_operato]~<->~%
\lstinline[style=tlsf_default]~ LOCKED);~\\%
\lstinline[style=tlsf_default]~  }~\\%
\lstinline[style=tlsf_default]~}~
\end{minipage}
\end{lrbox}

\maketitle

\begin{abstract}
  We present the Temporal Logic Synthesis Format (TLSF), a high-level
format to describe synthesis problems via Linear Temporal Logic
(LTL). The format builds upon standard LTL, but additionally allows to
use high-level constructs, such as sets and functions, to provide a
compact and human-readable representation. Furthermore, the format
allows to identify parameters of a specification such that a single
description can be used to define a family of problems. Additionally, we
present a tool to automatically translate the format into plain LTL,
which then can be used for synthesis by a solver. The tool also allows
to adjust parameters of the specification and to apply standard
transformations on the resulting formula.

\end{abstract}

\section{Introduction}
\label{sec:intro}
The automatic synthesis of reactive systems from formal specifications has been
one of the major challenges of computer science, and an active field 
of research, since the definition of the problem by Church~\cite{Church62}. For specifications in linear temporal logic the problem is 2EXPTIME-complete, and a number of fundamental approaches to solve this problem have 
been proposed~\cite{BL69,Rabin69,PnueliR89}, based on a translation of the specification into a game or an automaton. Recently, there has been a lot of work on solving synthesis problems more efficiently, either by restricting the specification language~\cite{BloemJPPS12,MorgensternS11}, or by a smart exploration of the search
space~\cite{Finkbeiner13,Ehlers12,fjr11,SohailS13,FinkbeinerJ12,FiliotJR13}. 

However, as already noted by Ehlers~\cite{Ehlers11a}, it has been very hard to compare different synthesis tools. A major reason for this was the lack of a common language and a benchmark library on which to compare tools. As a consequence, there has also been a lack of incentive for the development of efficient implementations of new synthesis approaches. 

To some extent, this has changed with the advent of the reactive synthesis competition (\syntcomp)~\cite{JacobsETAL16,JacobsETAL16a}, which has been organized in order to encourage the development of mature and efficient synthesis tools. However, \syntcomp thus far was restricted to safety specifications in an extension of the AIGER format~\cite{SYNTCOMP-format}, a low-level format that is not suited for writing expressive specifications by hand. Moreover, AIGER files directly represent a (safety) game, and the translation of a temporal logic specification to a suitable game (or other intermediate representations) is a non-trivial part of the synthesis problem that is removed from the picture if we start from an AIGER specification. 

In this paper, we introduce the \emph{temporal logic synthesis format} (TLSF), a high-level format for the specification of synthesis problems. The goal of TLSF is to create a format that (i) makes it \emph{convenient to write expressive specifications} by hand, and at the same time (ii)  is \emph{easy to support by synthesis tools}. 

To achieve the first goal, TLSF allows to define synthesis problems with high-level temporal logic specifications, and supports a number of additional features. These include user-defined enumeration types and signal buses, function declarations (including recursion), and the definition of parameters that allow to easily define parameterized families of synthesis problems. 

To achieve the second goal, we define a \emph{basic format} that is essentially restricted to linear temporal logic (LTL) without these additional features, and we supply the \emph{Synthesis Format Conversion Tool} (SyFCo) that can compile arbitrary TLSF specifications into the basic format. Hence, for synthesis tools it is sufficient to support the much simpler basic format. Moreover, the SyFCo tool also supports a number of additional features, including the translation to some existing specification formats like Promela LTL~\cite{PromelaLTL} or PSL~\cite{EF2006}, and is easily extensible to other formats.

To demonstrate the features of our specification format, we provide a version of the AMBA arbiter specification in TLSF. In addition to this, a large number of existing benchmarks have already been converted to TLSF, and can be found in our publicly available repository~\cite{SyFCo}.

TLSF will be used as a high-level format in several new tracks of \syntcomp in 2016~\cite{JacobsB16,JacobsETAL16b}. The goal is to develop and maintain a standard format for synthesis from high-level temporal logic specifications, and to use our repository of benchmarks as a starting point for a growing benchmark library that will be part of \syntcomp. The design decisions that went into TLSF are inspired by the findings of Schirmer~\cite{THSEBSCH}, who compared existing synthesis formats and made a first proposal towards the goals stated above.

\paragraph{Overview.} 
We present the basic version of the Temporal Logic Synthesis Format (TLSF)
in \secref{sec:basicformat}. In
\secref{sec:semantics} we discuss the intended semantics of a
specification, defined in terms of different implementation models.
The full format is introduced in
\secref{sec:format}, followed by an illustration of its main
features on an example in \secref{sec:example}.  In
\secref{sec:tool}, we give an overview of the SyFCo Tool.
Finally, we discuss possible extensions of the format in
\secref{sec:extensions}.

\section{The Basic Format}
\label{sec:basicformat}
A specification in the basic format consists of an \lstinline!INFO! section and a
\lstinline!MAIN! section:
\begin{equation*}
  \tlsfsec{info} \tlsfsec{main} 
\end{equation*}

\subsection{The INFO Section}
\label{sec:basicinfo}

The \lstinline!INFO! section contains the meta data of the specification, like a
title and some description\footnote{We use colored verbatim font to
  identify the syntactic elements of the specification.}. Furthermore,
it defines the underlying semantics of the specification (Mealy or
Moore / standard or strict implication) and the target model of the
synthesized implementation. Detailed information about supported
semantics and targets can be found in \secref{sec:semantics}.
Finally, a comma separated list of tags can be specified to identify
features of the specification, e.g., the restriction to a specific
fragment of LTL. A \tlsfid{tag} can be any string literal and is not
restricted to any predefined keywords.

\goodbreak

\vspace{1em}

\noindent
\lstinline!  INFO {!\\%
\lstinline!    TITLE:       "!$ \tlsfid{some title} $\lstinline!"!\\%
\lstinline!    DESCRIPTION: "!$ \tlsfid{some description} $\lstinline!"!\\%
\lstinline!    SEMANTICS:   !$ \tlsfid{semantics} $\\%
\lstinline!    TARGET:      !$ \tlsfid{target} $\\%
\lstinline!    TAGS:        !$ \tlsfid{tag} $%
\lstinline!,! $ \tlsfid{tag} $\lstinline!,!$ \ \ldots $\\%
\lstinline!  }!

\subsection{The MAIN Section}

The specification is completed by the \lstinline!MAIN! section, which 
contains the
partitioning of input and output signals, followed by the main
specification. The specification itself is separated into assumptions on the 
environment and desired properties of the system, and can additionally be 
distinguished into initial (\lstinline!INITIALLY!/\lstinline!PRESET!), 
invariant (\lstinline!REQUIRE!/\lstinline!ASSERT!), and arbitrary (
\lstinline!ASSUME!/\lstinline!GUARANTEE!) properties\footnote{In TLSF 
v1.0~\cite{JacobsK16}, \lstinline!ASSERT! was called \lstinline!INVARIANTS!, \lstinline!ASSUME! was called 
\lstinline!ASSUMPTIONS!, and \lstinline!GUARANTEE! was called \lstinline!GUARANTEES! (and subsections \lstinline!INITIALLY!, 
\lstinline!PRESET!, and \lstinline!REQUIRE! did not exist). TLSF v1.1 still supports the old 
identifiers.}.  Multiple declarations and expressions need
to be separated by a '\lstinline!;!'.

\vspace{1em}

\noindent
\lstinline!  MAIN {!\\%
\lstinline!    INPUTS    { !%
$ ( \tlsfsec{boolean signal declaration} $\lstinline!;!$ )^{*} $\lstinline! }!\\%
\lstinline!    OUTPUTS   { !%
$ ( \tlsfsec{boolean signal declaration} $\lstinline!;!$ )^{*} $\lstinline! }!\\%
\lstinline!    INITIALLY { !%
$ ( \tlsfsec{basic LTL expression} $\lstinline!;!$ )^{*} $\lstinline! }!\\%
\lstinline!    PRESET    { !%
$ ( \tlsfsec{basic LTL expression} $\lstinline!;!$ )^{*} $\lstinline! }!\\%
\lstinline!    REQUIRE   { !%
$ ( \tlsfsec{basic LTL expression} $\lstinline!;!$ )^{*} $\lstinline! }!\\%
\lstinline!    ASSERT    { !%
$ ( \tlsfsec{basic LTL expression} $\lstinline!;!$ )^{*} $\lstinline! }!\\%
\lstinline!    ASSUME    { !%
$ ( \tlsfsec{basic LTL expression} $\lstinline!;!$ )^{*} $\lstinline! }!\\%
\lstinline!    GUARANTEE { !%
$ ( \tlsfsec{basic LTL expression} $\lstinline!;!$ )^{*} $\lstinline! }!\\%
\lstinline!  }!

\vspace{1em}

\noindent
All subsections except \lstinline!INPUTS! and \lstinline!OUTPUTS! are optional.

\subsection{Basic Expressions}

A basic expression~$ e $ is either a boolean signal or a basic LTL
expression. Each basic expression has a corresponding type that is
$ \signals $ for boolean signals and $ \temporals $ for LTL
expressions.  Basic expressions can be composed to larger expressions
using operators.  An overview over the different types of expressions
and operators is given below.

\subsubsection{Boolean Signal Declarations}
A signal identifier is represented by a string consisting of lowercase
and uppercase letters~\mbox{('\lstinline!a!'-'\lstinline!z!'},
\mbox{'\lstinline!A!'-'\lstinline!Z!')},
numbers~\mbox{('\lstinline!0!'-'\lstinline!9!')},
underscores~\mbox{('\lstinline!_!')}, primes~\mbox{('\lstinline!'!')},
and at-signs~\mbox{('\lstinline!@!')} and does not start with a number
or a prime. Additionally, keywords like \lstinline!X!, \lstinline!G!
or \lstinline!U!, as defined in the rest of this document, are
forbidden.
An identifier is declared as either an input or an output signal. We
denote the set of declared input signals as $ \inputs $ and the set of
declared output signals as $ \outputs $, where
$ \inputs \cap \outputs = \emptyset $.  Then, a boolean signal
declaration simply consists of a signal identifier \tlsfid{name} from
$ \inputs \cup \outputs $.

\subsubsection{Basic LTL Expressions}
A basic LTL expression conforms to the following grammar, including
truth values, signals, boolean operators and temporal operators. For
easy parsing of the basic format, we require fully parenthesized
expressions, as expressed by the first of the following lines:
\begin{eqnarray*}
  \varphi & \equiv & 
  \text{\lstinline|(|} \varphi' \text{\lstinline|)|} \\
  \varphi' & \equiv & 
  \text{\lstinline|true|} \sep
  \text{\lstinline!false!} \sep
  s \text{\ \ \ for } s \in \inputs \cup \outputs \sep 
  \\ & &
  \text{\lstinline|!|} \varphi \sep
  \varphi \ \text{\lstinline!&&!} \ \varphi \sep
  \varphi \ \text{\lstinline!||!} \ \varphi \sep
  \varphi \ \text{\lstinline!->!} \ \varphi \sep
  \varphi \ \text{\lstinline!<->!} \ \varphi 
  \\ & & 
  \text{\lstinline!X!} \ \varphi \sep
  \text{\lstinline!G!} \ \varphi \sep
  \text{\lstinline!F!} \ \varphi \sep
  \varphi \ \text{\lstinline!U!} \ \varphi \sep
  \varphi \ \text{\lstinline!R!} \ \varphi \sep
  \varphi \ \text{\lstinline!W!} \ \varphi
\end{eqnarray*}
Thus, a basic LTL expression is either 
true, false, or a signal, or composed from these atomic expressions 
with boolean operators (negation,
conjunction, disjunction, implication, equivalence) and temporal
operators (next, globally, eventually, until, release, weak until).
The semantics of the boolean operators are defined in the usual way, and the temporal operators are defined in Appendix~\ref{apx_ltl}.

\section{Targets and Semantics}
\label{sec:semantics}
\subsection{Targets}
The \lstinline!TARGET! of the specification defines the implementation model that
a solution should adhere to. Currently supported targets are Mealy
automata \mbox{(\lstinline!Mealy!)}, whose output depends on the
current state and input, and Moore automata
\mbox{(\lstinline!Moore!)}, whose output only depends on the current
state. The differentiation is necessary since realizability of a 
specification depends on the target system model. For example, every 
specification that is realizable
under Moore semantics is also realizable under Mealy semantics, but
not vice versa. A formal description
of both automata models can be found in Appendix~\ref{apx_memo}.

\subsection{Semantics}
The \lstinline!SEMANTICS! of the specification defines how the formula
was intended to be evaluated, which also depends on an
implementation model.  We currently support four different semantics:
standard Mealy semantics \mbox{(\lstinline!Mealy!)}, standard Moore
semantics \mbox{(\lstinline!Moore!)}, strict Mealy semantics
\mbox{(\lstinline!Mealy,!} \lstinline!Strict!), and strict Moore semantics
\mbox{(\lstinline!Moore,Strict!)}. 

In the following, consider a specification where \lstinline!INITIALLY! evaluates to the LTL formula $\theta_e$, \lstinline!PRESET! evaluates to $\theta_s$, \lstinline!REQUIRE! evaluates to $\psi_e$, \lstinline!ASSERT! evaluates to $\psi_s$, \lstinline!ASSUME! evaluates to $\varphi_e$, and \lstinline!GUARANTEE! evaluates to $\varphi_s$. For specification sections that are not present, the respective formula is interpreted as \lstinline!true!.

\subsubsection{Standard semantics}
If the semantics is (non-strict) \mbox{\lstinline!Mealy!} or 
\mbox{\lstinline!Moore!}, and the \lstinline!TARGET! coincides with the semantics system 
model, then the specification is interpreted as the formula
\begin{equation*}
  \theta_e \rightarrow \left( \theta_s \land (\LTLglobally \psi_{e} \land \varphi_{e} \rightarrow \LTLglobally \psi_{s} \wedge \varphi_{s}) \right)
\end{equation*}
in standard LTL semantics (see Appendix~\ref{apx_ltl}). Note that we require that the \lstinline!PRESET! property $\theta_s$ holds whenever the \lstinline!INITIALLY! condition $\theta_e$ holds, regardless of other environment assumptions.

\subsubsection{Strict semantics}
If the semantics is \mbox{\lstinline!Mealy,Strict!} or 
\mbox{\lstinline!Moore,Strict!}, and the \lstinline!TARGET! coincides with the semantics 
system model, then the specification is interpreted under strict implication 
semantics (as used in the synthesis of GR(1) specifications), which is equivalent to the formula
\begin{equation*}
  \theta_e \rightarrow \left( \theta_s \land (\psi_{s} \LTLweakuntil \neg \psi_{e}) \land (\LTLglobally \psi_{e} \land \varphi_{e} \rightarrow \varphi_{s}) \right)
\end{equation*}
in standard LTL semantics. In this case, we additionally require that the \lstinline!ASSERT! property $\psi_s$ needs to hold at least as long as the \lstinline!REQUIRE! condition $\psi_e$ holds. 

Note that this gives us an easy way to convert a specification with strict semantics into one with non-strict semantics. For details on strict implication semantics, see Klein and 
Pnueli~\cite{KleinP10}, as well as Bloem et al.~\cite{BloemJPPS12}, from which we also take our definition and interpretation of the GR(1) fragment.\footnote{Note that in the conversion of
  \cite{BloemJPPS12}, the formula is strengthened by
  adding the
  formula~$ \LTLglobally (\LTLpastglobally \psi_{e} \rightarrow
  \psi_{s})$,
  where $ \LTLpastglobally \varphi $ is a Past-LTL formula and denotes
  that $ \varphi $ holds everywhere in the past. However, it is easy
  to show that our definition of strict semantics matches the
  definition of \cite{BloemJPPS12}. We prefer this
  notion, since it avoids the introduction of Past-LTL.}

\subsubsection{Conversion between system models}
If the implementation model of the \lstinline!SEMANTICS! differs from the \lstinline!TARGET! of a 
specification, we use a simple conversion to get a specification that is 
realizable in the target system model iff the original specification is 
realizable in the original system model: a specification in 
Moore semantics can be converted into Mealy semantics by prefixing all 
occurrences of input atomic propositions with an additional 
$\LTLnext$-operator. Similarly, we can convert from Mealy semantics to Moore 
semantics by prefixing outputs with a $ \LTLnext $-operator.

\section{The Full Format}
\label{sec:format}
In the full format, a specification consists of three sections: the
\lstinline!INFO! section, the \lstinline!GLOBAL! section and the \lstinline!MAIN! section. The \lstinline!GLOBAL!
section is optional.
\begin{equation*}
  \tlsfsec{info} [\tlsfsec{global}] \tlsfsec{main}
\end{equation*}
The \lstinline!INFO! section is the same as in the basic format, defined in
\secref{sec:basicinfo}. The \lstinline!GLOBAL! section can be used to define
parameters, and to bind identifiers to expressions that can be used
later in the specification. The \lstinline!MAIN! section is used as before, but
can use extended sets of declarations and expressions.

We define the \lstinline!GLOBAL! section in \secref{sec:global}, and the changes
to the \lstinline!MAIN! section compared to the basic format in
\secref{sec:main-full}. The extended set of expressions that can be
used in the full format is introduced in \secref{sec:expressions},
enumerations, extended signal and function declarations in
\secref{sec:enumerations} and \ref{sec:functions}, and additional
notation in \secref{sec:bigoperator}--\ref{sec:comments}.

\subsection{The GLOBAL Section}
\label{sec:global}

The \lstinline!GLOBAL! section consists of the \lstinline!PARAMETERS!
subsection, defining the identifiers that parameterize the specification, 
and the \lstinline!DEFINITIONS!
subsection, that allows to define functions, enumerations and to bind
identifiers to complex expressions. Multiple declarations need to be
separated by a '\lstinline!;!'. The section and its subsections are
optional.

\vspace{1em}

\noindent
\lstinline!  GLOBAL {!\\%
\lstinline!    PARAMETERS { !\\%
\lstinline!      !$ ( \tlsfid{identifier} $\lstinline! = !%
$ \tlsfsec{numerical expression} $\lstinline!;!$ )^{*} $\\%
\lstinline!    }!\\%
\lstinline!    DEFINITIONS  { !\\%
\lstinline!      !%
$ ((\tlsfid{function declaration} \sep \tlsfid{enum declaration} 
\sep \tlsfid{identifier} $%
\lstinline! = !$ \tlsfsec{expression}) $\lstinline!;!$ )^{*} $\\%
\lstinline!    }!\\%
\lstinline!  }!

\subsection{The MAIN Section}
\label{sec:main-full}

Like in the basic format, the \lstinline!MAIN! section contains the partitioning
of input and output signals, as well as the main
specification. However, signal declarations can now contain signal
buses, and LTL expressions can use parameters, functions, and
identifiers defined in the \lstinline!GLOBAL! section.

\vspace{1em}

\noindent
\lstinline!  MAIN {!\\%
\lstinline!    INPUTS    { !%
$ ( \tlsfsec{signal declaration} $\lstinline!;!$ )^{*} $\lstinline! }!\\%
\lstinline!    OUTPUTS   { !%
$ ( \tlsfsec{signal declaration} $\lstinline!;!$ )^{*} $\lstinline! }!\\%
\lstinline!    INITIALLY { !%
$ ( \tlsfsec{LTL expression} $\lstinline!;!$ )^{*} $\lstinline! }!\\%
\lstinline!    PRESET    { !%
$ ( \tlsfsec{LTL expression} $\lstinline!;!$ )^{*} $\lstinline! }!\\%
\lstinline!    REQUIRE   { !%
$ ( \tlsfsec{LTL expression} $\lstinline!;!$ )^{*} $\lstinline! }!\\%
\lstinline!    ASSERT    { !%
$ ( \tlsfsec{LTL expression} $\lstinline!;!$ )^{*} $\lstinline! }!\\%
\lstinline!    ASSUME    { !%
$ ( \tlsfsec{LTL expression} $\lstinline!;!$ )^{*} $\lstinline! }!\\%
\lstinline!    GUARANTEE { !%
$ ( \tlsfsec{LTL expression} $\lstinline!;!$ )^{*} $\lstinline! }!\\%
\lstinline!  }!

\vspace{1em}

\noindent
As before, all subsections except \lstinline!INPUTS! and \lstinline!OUTPUTS! are
optional.

\subsection{Expressions}
\label{sec:expressions}

An expression~$ e $ is either a boolean signal, an $ n $-ary signal
(called bus), an enumeration type, a numerical expression, a boolean expression, an LTL
expression, or a set expression. Each expression has a corresponding
type that is either one of the basic types:
$ \signals, \buses, \enums, \nats, \bools, \temporals $, or a recursively
defined set type~$ \mathcal{S}_{\arbitrary} $ for some type~$ \arbitrary $.

As before, an identifier is represented by a string consisting of
lowercase and uppercase
letters~(\mbox{'\lstinline!a!'-'\lstinline!z!'},
\mbox{'\lstinline!A!'-'\lstinline!Z!'}),
numbers~('\lstinline!0!'-'\lstinline!9!'),
underscores~('\lstinline!_!'), primes~('\lstinline!'!'), and
at-signs~('\lstinline!@!') and does not start with a number or a
prime.  In the full format, identifiers are bound to expressions of
different type. We denote the respective sets of identifiers by
$ \signalids $, $ \busids $, $ \enumids $, $ \natids $, $ \boolids $,
$ \temporalids $, and $ \atypeids $.  Finally, basic expressions can
be composed to larger expressions using operators. In the full format,
we do not require fully parenthesized expressions. If an expression is
not fully parenthesized, we use the precedence order given in
Table~\ref{tab:precedence}. An overview over the all types of
expressions and operators is given below.

\begin{table}
  \centering

  \renewcommand{\arraystretch}{1.05}

  \begin{tabular}{|c|l|l|c|c|}
    \hline 
    \textbf{Precedence} & \textbf{Operator} & \textbf{Description} & \textbf{Arity} &  \textbf{Associativity} \\
    \hline
    \hline
    \multirow{6}{*}{1} & \lstinline!+[!$ \cdot $\lstinline!]!\ (\lstinline!SUM[!$ \cdot $\lstinline!]!) & sum & \multirow{6}{*}{unary} & \\
      & \lstinline!*[!$ \cdot $\lstinline!]!\ (\lstinline!PROD[!$ \cdot $\lstinline!]!) & product & & \\
      & \lstinline!|!$ \cdots $\lstinline!|!\ (\lstinline!SIZE!) & size & & \\
      & \lstinline!MIN! & minimum & & \\
      & \lstinline!MAX! & maximum & & \\
      & \lstinline!SIZEOF! & size of a bus & & \\
    \hline
    2 & \lstinline!*!\ (\lstinline!MUL!) & multiplication & binary & left-to-right \\
    \hline
    \multirow{2}{*}{3} & \lstinline!/!\ (\lstinline!DIV!) & integer division & \multirow{2}{*}{binary} & \multirow{2}{*}{right-to-left} \\
      & \lstinline!%!\ (\lstinline!MOD!) & modulo & & \\
    \hline
    \multirow{2}{*}{4} & \lstinline!+!\ (\lstinline!PLUS!) & addition & \multirow{2}{*}{binary} & \multirow{2}{*}{left-to-right} \\
      & \lstinline!-!\ (\lstinline!MINUS!) & difference & & \\
    \hline 
    \multirow{2}{*}{5} & \lstinline!(*)[!$ \cdot $\lstinline!]!\ (\lstinline!CAP[!$ \cdot $\lstinline!]!) & intersection & \multirow{2}{*}{unary} & \\
      & \lstinline!(+)[!$ \cdot $\lstinline!]!\ (\lstinline!CUP[!$ \cdot $\lstinline!]!) & union &  & \\
    \hline
    6 & \lstinline!(\)!\ (\lstinline!(-)!,\lstinline!SETMINUS!) & set difference & binary & right-to-left \\
    \hline
    7 & \lstinline!(*)!\ (\lstinline!CAP!) & intersection & binary & left-to-right \\
    \hline
    8 & \lstinline!(+)! (\lstinline!CUP!) & union & binary & left-to-right \\
    \hline
    \multirow{6}{*}{9} & \lstinline!==!\ (\lstinline!EQ!) & equality & \multirow{6}{*}{binary} & \multirow{6}{*}{left-to-right} \\
      & \lstinline~!=~\ (\lstinline!/=!, \lstinline!NEQ!) & inequality & & \\
      & \lstinline!<!\ (\lstinline!LE!) & smaller than & & \\
      & \lstinline!<=!\ (\lstinline!LEQ!) & smaller or equal than & & \\
      & \lstinline!>!\ (\lstinline!GE!) & greater then & & \\
      & \lstinline!>=!\ (\lstinline!GEG!) & greater or equal than & & \\
    \hline
    10 & \lstinline!IN!\ (\lstinline!ELEM!, \lstinline!<-!) & membership & binary & left-to-right \\
    \hline
    \multirow{6}{*}{11} & \lstinline~!~\ (\lstinline!NOT!) & negation & \multirow{6}{*}{unary} & \\
       & \lstinline!X! & next & & \\
       & \lstinline!F! & finally & & \\
       & \lstinline!G! & globally & & \\
       & \lstinline!&&[!$ \cdot $\lstinline!]!\ (\lstinline!AND[!$ \cdot $\lstinline!]!, \lstinline!FORALL[!$ \cdot $\verb!]!) & conjunction & & \\
       & \lstinline!||[!$ \cdot $\lstinline!]!\ (\lstinline!OR[!$ \cdot $\lstinline!]!, \lstinline!EXISTS[!$ \cdot $\verb!]!) & disjunction & & \\
    \hline
    12 & \lstinline!&&!\ (\lstinline!AND!) & conjunction & binary & left-to-right \\
    \hline
    13 & \lstinline!||!\ (\lstinline!OR!) & disjunction & binary & left-to-right \\     
    \hline
    \multirow{2}{*}{14} & \lstinline!->!\ (\lstinline!IMPLIES!) & implication & \multirow{2}{*}{binary} & \multirow{2}{*}{right-to-left} \\   
       & \lstinline!<->!\ (\lstinline!EQUIV!) & equivalence &  & \\     
    \hline
    15 & \lstinline!W! & weak until & binary & right-to-left \\
    \hline
    16 & \lstinline!U! & until & binary & right-to-left \\
    \hline
    17 & \lstinline!R! & release & binary & left-to-right \\
    \hline
    18 & \!{\color{blue!70!black}\raisebox{-9pt}{\scalebox{1.7}{\textasciitilde}}} \vspace{-5pt} & pattern match & binary & left-to-right \\
    \hline
    19 & \lstinline!:! & guard & binary & left-to-right \\    
    \hline
  \end{tabular}

  \renewcommand{\arraystretch}{1}

  \caption{The table lists the precedence, arity and associativity of
    all expression operators. Also consider the alternative names in
    brackets which can be used instead of the symbolic
    representations.}
  \label{tab:precedence}
\end{table}

\subsubsection{Numerical Expressions}
A numerical expression~$ e_{\nats} $ conforms to the following
grammar:
\begin{eqnarray*}
  e_{\nats} & \equiv &
  i \text{\ \ \ for } i \in \natids \sep 
  n  \text{\ \ \ for } n \in \nats \sep
  e_{\nats} \ \text{\lstinline!+!} \ e_{\nats} \sep
  e_{\nats} \ \text{\lstinline!-!} \ e_{\nats} \sep
  e_{\nats} \ \text{\lstinline!*!} \ e_{\nats} \sep
  e_{\nats} \ \text{\lstinline!/!} \ e_{\nats} \sep
  e_{\nats} \ \, \text{\lstinline!\%!} \ \, e_{\nats} \\ 
  & &
  \text{\lstinline!|!} e_{\mathcal{S}_{\arbitrary}} \text{\lstinline!|!} \sep
  \text{\lstinline!MIN!} \ e_{\mathcal{S}_{\nats}} \sep
  \text{\lstinline!MAX!} \ e_{\mathcal{S}_{\nats}} \sep
  \text{\lstinline!SIZEOF!} \ s \ \ \text{ for } s \in \busids
\end{eqnarray*}
Thus, a numerical expression either represents an identifier (bound to
a numerical value), a numerical constant, an addition, a subtraction, a
multiplication, an integer division, a modulo operation, the size of a
set, the minimal/maximal value of a set of naturals, or the size (i.e., width) of a
bus, respectively. The semantics are defined in the usual way.

\subsubsection{Set Expressions}
A set expression~$ e_{\mathcal{S}_{\arbitrary}} $, containing elements
of type $ \mathbb{X} $, conforms to the following grammar:
\begin{eqnarray*}
  e_{\mathcal{S}_{\arbitrary}} & \equiv &
  i  \text{\ \ \ for } i \in \atypeids \sep
  \text{\lstinline!\{!} \, e_{\arbitrary} \text{\lstinline!,!} \, e_{\arbitrary}
  \text{\lstinline!,!} \ldots \text{\lstinline!,!} \, e_{\arbitrary} \,
  \text{\lstinline!\}!} \sep
  \text{\lstinline!\{!} \, e_{\nats} \text{\lstinline!,!} \, e_{\nats} \, 
  \text{\lstinline!..!}\,  e_{\nats} \, \text{\lstinline!\}!} \sep 
  \\ &&
  e_{\mathcal{S}_{\arbitrary}} \, \text{\lstinline!(+)!} \ e_{\mathcal{S}_{\arbitrary}} \sep
  e_{\mathcal{S}_{\arbitrary}} \, \text{\lstinline!(*)!} \ e_{\mathcal{S}_{\arbitrary}} \sep
  e_{\mathcal{S}_{\arbitrary}} \, \text{\lstinline!(\\)!} \ e_{\mathcal{S}_{\arbitrary}}
\end{eqnarray*}
Thus, the expression~$ e_{\mathcal{S}_{\arbitrary}} $ either
represents an identifier (bound to a set of values of type
$ \arbitrary $), an explicit list of elements of type $ \arbitrary $,
a list of elements specified by a range (for $ \arbitrary = \nats $),
a union of two sets, an intersection or a difference,
respectively. The semantics of a range expression
\lstinline!{!$ x $\lstinline!,!$ y $\lstinline!..!$ z $\lstinline!}!
are defined for $ x < y $ via:
\begin{equation*}
  \{ n \in \nats \mid x \leq n \leq z \wedge \exists j.\ n = x +
  j \cdot (y-x) \}.
\end{equation*}
The semantics of all other expressions are defined as usual. Sets
contain either positive integers, boolean expressions, LTL
expressions, buses, signals, or other sets of a specific type.

\subsubsection{Boolean Expressions}
A boolean expression~$ e_{\bools} $ conforms to the following
grammar:
\begin{eqnarray*}
  e_{\bools} & \equiv & 
  i  \text{\ \ \ for } i \in \boolids \sep
  e_{\arbitrary} \ \text{\lstinline!IN!} \ e_{\mathcal{S}_{\arbitrary}} \sep
  \text{\lstinline|true|} \sep
  \text{\lstinline!false!} \sep
  \text{\lstinline|!|}\,e_{\bools} \sep
  \\ & & 
  e_{\bools} \ \text{\lstinline!&&!} \ e_{\bools} \sep
  e_{\bools} \ \text{\lstinline!||!} \ e_{\bools} \sep
  e_{\bools} \ \text{\lstinline!->!} \ e_{\bools} \sep
  e_{\bools} \ \text{\lstinline!<->!} \ e_{\bools} \sep
  \\ & & 
  e_{\nats} \ \text{\lstinline!==!} \ e_{\nats} \sep
  e_{\nats} \ \text{\lstinline~!=~} \ e_{\nats} \sep
  e_{\nats} \ \text{\lstinline!<!} \ e_{\nats} \sep
  e_{\nats} \ \text{\lstinline!<=!} \ e_{\nats} \sep
  e_{\nats} \ \text{\lstinline!>!} \ e_{\nats} \sep
  e_{\nats} \ \text{\lstinline!>=!} \ e_{\nats} 
\end{eqnarray*}
Thus, a boolean expression either represents an identifier (bound to a
boolean value), a membership test, true, false, a negation, a
conjunction, a disjunction, an implication, an equivalence, or an
equation between two positive integers (equality, inequality, less
than, less or equal than, greater than, greater or equal than),
respectively. The semantics are defined in the usual way. Note that
signals are not allowed in a boolean expression, but only in an LTL
expression.

\subsubsection{LTL Expressions}
An LTL expression~$ \varphi $ conforms to the same grammar as a
boolean expression, except that it additionally includes signals and
temporal operators.
%
\begin{eqnarray*}
  \varphi & \equiv & \ldots \sep
  i \text{\ \ for } i \in \temporalids \sep
  s \text{\ \ for } s \in \signalids \sep
  b \text{\lstinline![!} e_{\nats} 
  \text{\lstinline!]!} \text{ for } b 
  \in \busids \sep
  \\ & & 
  b_{0} \ \, \text{\lstinline!==!} \ \, b_{1} \text{\ \ \ for } b_{j} \in \busids \text{ and } b_{1-j} \in \enumids \sep
  b_{0} \ \, \text{\lstinline~!=~} \ \, b_{1} \text{\ \ \ for } b_{j} \in \busids \text{ and } b_{1-j} \in \enumids \sep
  \\ & &
  \text{\lstinline!X!} \ \varphi \sep
  \text{\lstinline!G!} \ \varphi \sep
  \text{\lstinline!F!} \ \varphi \sep
  \varphi \ \text{\lstinline!U!} \ \varphi \sep
  \varphi \ \text{\lstinline!R!} \ \varphi \sep
  \varphi \ \text{\lstinline!W!} \ \varphi 
\end{eqnarray*}
\goodbreak
\noindent
Thus, an LTL expression additionally can represent an identifier bound
to an LTL formula, a signal, an $e_{\nats} $-th signal of a bus, a
next operation, a restriction of a bus to a set of enumeration
valuations via equality or inequality, a globally operation, an
eventually operation, an until operation, a release operation, or a
weak until operation, respectively. Note that every boolean expression
is also an LTL expression, thus we allow the use of identifiers that
are bound to boolean expressions as well. A formal definition of the
semantics of the temporal operators is given in
Appendix~\ref{apx_ltl}. The semantics of expressions involving bus
operations is defined in the subsequent sections.

\subsection{Enumerations}
\label{sec:enumerations}

An enumeration declaration conforms to the following grammar:
\begin{equation*}
  \text{\lstinline!enum!} \ \tlsfid{enumtype} \ \text{\lstinline!=!} \ 
  \Big( \tlsfid{identifier} \ \text{\lstinline!:!} \ 
  (\text{\lstinline!0!} \!\! \sep \!\! \text{\lstinline!1!} \!\! \sep 
  \!\! \text{\lstinline!*!})^{n} \big( \text{\lstinline!,!} \ 
  (\text{\lstinline!0!} \!\! \sep \!\! \text{\lstinline!1!} \!\! \sep
  \!\! \text{\lstinline!*!})^{n} \big)^{*} \Big)^{+}
\end{equation*}
for some arbitrary but fix positive integer $ n > 0 $. As an example consider the
enumeration \lstinline!Positions!, which declares the enumeration
identifiers \lstinline!LEFT!, \lstinline!MIDDLE!, \lstinline!RIGHT!, and
\lstinline!UNDEF! as members of $ \enumids $:

\vspace{0.6em}

\noindent
\lstinline!  enum Position =!\\%
\lstinline!    LEFT:   100!\\%
\lstinline!    MIDDLE: 010!\\%
\lstinline!    RIGHT:  001!\\%
\lstinline!    UNDEF:  11*, 1*1, *11!\\
   
\vspace{0.6em}

\noindent We use \lstinline!0! to identify the absent signal,
\lstinline!1! to identify the present signal and \lstinline!*! for
either of both. Each identifier then refers to at least one concrete
signal valuation sequence. Multiple values can be denoted by sequences
with a \lstinline!*!, as well as by comma separated
lists. Furthermore, the identifier of each declared valuation has to
be unique. Not all possible valuations have to be identified.

Enumeration identifiers can only be used in comparisons against buses
inside an LTL expression, where we require that the corresponding bus
has the same width as the valuation compared to. It defines a boolean
constraint on the bus, restricting it to the different valuations,
bound to the identifier, e.g., the expressions %
\lstinline!b == RIGHT! and %
\lstinline~!b[0] && !b[1] && b[2]~ are semantically equivalent, as
well as %
\lstinline!b /= UNDEF! and %
\lstinline~!((b[0] && b[1]) || (b[0] && b[2]) || (b[1] && b[2]))~.

\subsection{Signals and Buses}
\label{sec:signals}

A single signal declaration consists of the name of the signal.  As
for the basic format, signals are declared as either input or output
signals, denoted by $ \inputs $ and $ \outputs $, respectively. A bus
declaration additionally specifies a signal width, i.e., a bus
represents a finite set of signals. The signal width is either given
by a numerical value or via an enumeration type.
\begin{equation*}
  \tlsfid{name} \sep
  \tlsfid{name} \text{\lstinline![!} 
  e_{\nats} \text{\lstinline!]!} \sep 
  \tlsfid{enumtype} \tlsfid{name}
\end{equation*}
Semantically, a signal declaration \lstinline!s! specifies a signal
$ s \in \inputs \cup \outputs $, where a bus declaration
\lstinline!b[n]! specifies $ n $ signals~\lstinline!b[0]!,
\lstinline!b[1]!, $ \ldots $, \lstinline!b[n-1]!, with either
\lstinline!b[i]!$ \in \inputs $ for all $ 0 \leq i < n $, or
\lstinline!b[i]!$ \in \outputs $ for all $ 0 \leq i < n $. A bus
specified via an enumeration type has the same width as the valuations
of the corresponding enumeration.

Buses which are declared using an enumeration type, where not all
valuations are related to an identifier\footnote{See e.g.\ the
  \lstinline!000!  valuation of the example of \secref{sec:signals}}
induce an implicit constraint on the corresponding signals: if
the bus corresponds to a set of input signals, then the global
requirement that no other than the defined valuations appear on this
bus is imposed. If it corresponds to a set of output signals, then the
equivalent global invariant is imposed.

Finally, note that we use \lstinline!b[i]! to access the $ i $-th
value of $ b $, i.e., we use the same syntax as for the declaration
itself\hspace{1pt}\footnote{C-Array Syntax Style}. Also note that
for the declared signals~$ s $, we have
$ s \in \inputs \cup \outputs \subseteq \signalids $, and for the
declared buses $ b $, we have $ b \in \busids $.

\subsection{Function Declarations}
\label{sec:functions}

As another feature, one can declare (recursive) functions of arbitrary
arity inside the \lstinline!DEFINITIONS! section. Functions can be used to define
simple macros, but also to generate complex formulas from a given set
of parameters. A declaration of a function of arity~$ n $ has the form
%
\begin{equation*}
  \tlsfid{function name} \text{\lstinline!(!}
  \tlsfid{arg\ensuremath{_{1}}} \text{\lstinline!,!} 
  \tlsfid{arg\ensuremath{_{2}}} \text{\lstinline!,!}  
  \ldots \text{\lstinline!,!}
  \tlsfid{arg\ensuremath{_{n}}} \text{\lstinline!) =\ !}
  (e_{c})^{+},
\end{equation*}
where
$ \tlsfid{arg\ensuremath{_{1}}}, \tlsfid{arg\ensuremath{_{2}}},
\ldots, \tlsfid{arg\ensuremath{_{n}}} $
are fresh identifiers that can only be used inside the
sub-expressions~$ e_{c} $. An expression~$ e_{c} $ conforms to the
following grammar:
\begin{equation*}
  e_{c} \ \, \equiv \ \, e \sep
  e_{\bools} \ \text{\lstinline!:!} \ e \sep
  e_{\mathbb{P}} \ \text{\lstinline!:!} \ e 
  \qquad \qquad \text{where } 
  \ \ e \ \, \equiv  \ \,
  e_{\nats} \sep
  e_{\bools} \sep
  e_{\mathcal{S}_{\arbitrary}} \sep
  \varphi
\end{equation*}
Thus, a function can be bound to any expression~$ e $, parameterized
in its arguments, which additionally may be guarded by some boolean
expression~$ e_{\bools} $, or a pattern match~$ e_{\pats} $. If the
regular expression~$ (e_{c})^{+} $ consists of more than one
expression~$ e_{c} $, then the function binds to the first expression
whose guard evaluates to \lstinline!true! (in the order of their
declaration). Furthermore, the special guard \lstinline!otherwise! can
be used, which evaluates to \lstinline!true! if and only if all other
guards evaluate to \lstinline!false!. Expressions without a guard are
implicitly guarded by \lstinline!true!. All sub-expressions~$ e_{c} $
need to have the same type $ \arbitrary $. For every instantiation of
a function by given parameters, we view the resulting expression
$ e_{\arbitrary} $ as an identifier in~$ \Gamma_{\arbitrary} $, bound
to the result of the function application.
%
 
\subsubsection{Pattern Matching}
\label{sec_patterns}
Pattern matches are special guards of the form
\begin{equation*}
  e_{\pats} \equiv \ \, \varphi \ \lstilde \ \varphi' ,
  \vspace{-0.5em}
\end{equation*}
which can be used to describe different behavior depending on the
structure of an LTL expression. Hence, a guard~$ e_{\pats} $ evaluates
to \lstinline!true! if and only if $ \varphi $ and $ \varphi' $ are
structurally equivalent, with respect to their boolean and temporal
connectives. Furthermore, identifier names that are used in
$ \varphi' $ need to be fresh, since every identifier expression that
appears in $ \varphi' $ is bound to the equivalent sub-expression in
$ \varphi $, which is only visible inside the right-hand-side of the
guard. Furthermore, to improve readability, the special
identifier~\lstinline!_! (wildcard) can be used,
which always remains unbound. To clarify this feature, consider the
following function declaration:

\vspace{0.6em}

\noindent
\lstinline!  fun(f) =!\\%
\vspace{-0.4em}%
\lstinline!    f !$ \hspace{-1.5pt} \lstilde \hspace{-1.5pt} $%
\lstinline! a U _: a!\\%
\lstinline!    otherwise: X f!
   
\vspace{0.6em}

\noindent The function $ \textit{fun} $ gets an LTL formula $ f $ as a
parameter. If $ f $ is an until formula of the form
$ \varphi_{1} \LTLuntil \varphi_{2} $, then $ \textit{fun}(f) $ binds
to $ \varphi_{1} $, otherwise $ \textit{fun}(f) $ binds to
$ \LTLnext f $.

\subsection{Big Operator Notation}
\label{sec:bigoperator}

It is often useful to express parameterized expressions using ``big''
operators, e.g., we use $ \Sigma $ to denote a sum over multiple
sub-expressions, $ \Pi $ to denote a product, or $ \bigcup $ to denote
a union. It is also possible to use this kind of notion in this
specification format. The corresponding syntax looks as follows:
\begin{equation*}
  \tlsfid{op} \text{\lstinline![!} \, 
  \tlsfid{id\ensuremath{_{0}}} \, 
  \text{\lstinline!IN!} \; e_{\mathcal{S}_{\arbitrary_{0}}} \! 
  \text{\lstinline!,!}\, \tlsfid{id\ensuremath{_{1}}} \, 
  \text{\lstinline!IN!} \; e_{\mathcal{S}_{\arbitrary_{1}}} \! 
  \text{\lstinline!,!}\, \ldots \, 
  \text{\lstinline!,!} \,\tlsfid{id\ensuremath{_{n}}} \, 
  \text{\lstinline!IN!} \; e_{\mathcal{S}_{\arbitrary_{n}}} 
  \text{\lstinline!]!} \, e_{\arbitrary} 
\end{equation*}
Let $ x_{j} $ be the identifier represented by
$ \tlsfid{id\ensuremath{_{j}}} $ and $ S_{j} $ be the set represented by
$ e_{\mathcal{S}_{\arbitrary_{j}}}\! $. Further, let
$ \bigoplus $ be the mathematical operator corresponding to
$ \tlsfid{op} $. Then, the above expression corresponds to the
mathematical expression:
\begin{equation*}
  \bigoplus\limits_{x_{0} \in S_{0}} \ 
  \bigoplus\limits_{x_{1} \in S_{1}} \ \cdots \ 
  \bigoplus\limits_{x_{n} \in S_{n}}
  \big( e_{\arbitrary} )
\end{equation*}
Note that $ \tlsfid{id\ensuremath{_{0}}} $ is already bound in
expression $ e_{\mathcal{S}_{\arbitrary_{1}}} \!$,
$ \tlsfid{id\ensuremath{_{1}}} $ is bound in
$ e_{\mathcal{S}_{\arbitrary_{2}}} \!$, and so forth. The syntax is
supported by every operator
$ \tlsfid{op} \in \{ \text{\lstinline!+!},
\text{\lstinline!*!}, \text{\lstinline!(+)!},
\text{\lstinline!(*)!}, \text{\lstinline!&&!},
\text{\lstinline!||!} \} $.

\subsection{Syntactic Sugar}
\label{sec:syntacticsugar}

To improve readability, there is additional syntactic sugar, which can
be used beside the standard syntax. Let $ n $ and $ m $ be numerical
expressions, then

\begin{itemize}

\item \lstinline!X[!$ n $\lstinline!]!$ \; \varphi $ denotes a stack
  of $ n $ next operations, e.g.: \\[0.5em]
  \lstinline!  X[3] a!$ \ \, \equiv \ \, $ \lstinline!X X X a!

\item \lstinline!F[!$ n $\lstinline!:!$ m $\lstinline!]!$ \; \varphi $
  denotes that $ \varphi $ holds somewhere between the next $ n $ and
  $ m $ steps, e.g.: \\[0.5em]
  \lstinline!  F[2:3] a!$ \ \, \equiv \ \, $\lstinline!X X(a || X a)!

\item \lstinline!G[!$ n $\lstinline!:!$ m $\lstinline!]!$ \; \varphi $
  denotes that $ \varphi $ holds everywhere between the next $ n $ and
  $ m $ steps, e.g.: \\[0.5em]
  \lstinline!  G[1:3] a!$ \ \, \equiv \ \, $\lstinline!X(a && X(a && X a))!

\item $ \tlsfid{op} $\lstinline![!$ \, \ldots $\lstinline!,!$ \, n \, \circ_{1} 
  \tlsfid{id} \circ_{2} \, m \, $\lstinline!,!$ \ldots $\lstinline!]!$ \, e_{X} $
  denotes a big operator application, where  $ n \, \circ_{1} \tlsfid{id} 
  \circ_{2} \, m $ with $ \circ_{1}, \circ_{2} \in \{ \text{\lstinline!<!},
  \text{\lstinline!<=!} \} $ denotes that \tlsfid{id} ranges from $ n $ to
  $ m $. The inclusion of $ n $ and $ m $ depends on the
  choice of $ \circ_{1} $ and $ \circ_{2} $, respectively. Thus, the
  notation provides an alternative to membership in combination with
  set ranges, e.g.: \\[0.5em]
  \lstinline!  &&[0 <= i < n] b[i]!$ \ \, \equiv \ \, 
  $\lstinline!&&[i IN {0,1..n-1}] b[i]!
  %

\end{itemize}

\subsection{Comments}
\label{sec:comments}

It is possible to use C style comments anywhere in the specification,
i.e., there are single line comments initialized by \lstinline!//! and
multi line comments between \lstinline!/*! and
\lstinline!*/!.  Multi line comments can be nested.

\section{Example: A Decomposed AMBA Arbiter}
\label{sec:example}
To get a feeling for the interplay of the aforementioned features,
we present a specification of an arbiter for 
\textsc{Arm}'s \textit{Advanced
  Microcontroller Bus Architecture} (AMBA)~\cite{Amba:1999} in TLSF,
decomposed into multiple components as depicted in Figure~\ref{fig:amba}. 
Inputs of the system are requests (HBUSREQ) from masters that want 
to access the bus, and a ready signal (HREADY) from the clients that the 
masters want to talk to. Additionally, each master has a signal for locking the bus (HLOCK), and different types of locked accesses can be requested (via HBURST). The main output of the system is the number of the master that currently owns the bus (HMASTER, in a binary encoding), and a signal for whether the bus is locked (HMASTLOCK). Additionally, there are outputs for the next master that will get access to the bus (HGRANT, unary encoding).

Our encoding
is inspired by existing encodings~\cite{Jobstmann:2007}, but also 
includes some new
design aspects with respect to the decomposition. We only consider the
TLSF encoding of the components \textsc{decode}, \textsc{encode} and
\textsc{arbiter} in detail. Encodings of the remaining components
can be found in App.~\ref{apx:components}.

First, consider the \textsc{decode} component, whose encoding is
depicted in Figure~\ref{fig:decode}. The component reads the different
values of the HBURST bus and splits them up into separate, mutually
exclusive signals. Clearly, enumerations are perfectly suited to
describe such a behavior. 

\begin{figure}[H]
  \centering
  \scalebox{0.69}{\usebox{\architecture}}
\caption{Decomposition of the AMBA AHB arbiter.}
\label{fig:amba}
\end{figure}
Next, consider the \textsc{arbiter} component, granting the bus to the
different masters. The component selects a new master, whenever all
other components completed their tasks (signaled by
\textsc{ALLREADY}). Furthermore, every master requesting the bus is
eventually granted access to it, where every request needs the be held
until the access is granted. Additionally, we require that every
assignment of a new master triggers the \textsc{DECIDE} flag, which
has to be raised one time step in advance to inform other components
early about the change. Finally, the request signal of the granted bus
is mirrored by \linebreak \vspace{-2em}
\begin{figure}[H]
  \centering
  \begin{tikzpicture}
  \node[anchor=north] at (-0.05,0) {\usebox{\cArbiter}};
  \node[anchor=north] at (7.85,0) {\usebox{\cEncode}};
  \draw (3.9,0) -- (3.9,-19.9);
  \draw (-4,-19.9) rectangle (11.49,0);
  \end{tikzpicture}
  \vspace{-1em}
  \caption{The \textsc{arbiter} component (left) and the
    \textsc{encode} component (right) of the decomposed AMBA AHB
    arbiter.}
  \label{fig:arbitenc}  
\end{figure}
\noindent the \textsc{BUSREQ} output. The encoding of the
component is depicted in Figure~\ref{fig:arbitenc} on the left. It
uses straightforward formulations of the aforementioned properties in
TLSF, which integrate the behavior informally described above. Note
that the whole encoding is parameterized in the number of
masters~$ n $.
\begin{figure}[H]
  \centering
  \begin{tikzpicture}
    \node[anchor=north] (A) at (0,0) {\usebox{\cDecodeA}};
    \node[anchor=north] (B) at (7.5,0) {\usebox{\cDecodeB}};
    \draw (A.south west) rectangle (B.north east);
  \end{tikzpicture}
  \caption{The \textsc{decode} component of the decomposed AMBA AHB arbiter.}
  \label{fig:decode}
\end{figure}

As our final example, consider the encoding of the \textsc{encode}
component depicted on the right side of Figure~\ref{fig:arbitenc}. The
component identifies the master currently holding the bus via a binary
number, encoded logarithmically in the number of masters. Furthermore,
the component is only enabled as long as the \textsc{HREADY} input is
high.  We observe that the translation from unary to binary can be
easily described using a function mapping the unary values to the
corresponding binary ones. Inspecting the encoding shows that
the semantics of the function
are derivable straightforwardly from the declaration, due to the close
relation to the equivalent mathematical representation.

The remaining components are included in Appendix~\ref{apx:components}.

\section{The SyFCo Tool}
\label{sec:tool}
We created the Synthesis Format Conversion Tool (SyFCo)~\cite{SyFCo}
that can interpret the high level constructs of the format and
supports transformation of the specification to plain LTL. The
tool has been designed to be modular with respect to the supported
output formats and semantics. Furthermore, the tool can identify and
manipulate parameters, targets and semantics of a specification on the
fly, and thus allows comparative studies, as it is for example needed
in the reactive synthesis competition.

\vspace{1em}

\noindent The main features of the tool can be summarized as follows:

\begin{itemize}

\item Evaluation of high level constructs in the full format to reduce
  full TLSF to basic TLSF.

\item Transformation to other existing specification formats, like
  Promela LTL~\cite{PromelaLTL}, PSL~\cite{EF2006},
  Unbeast~\cite{E2010}, Wring~\cite{SomenziB00}, or SLUGS~\cite{ERF2013,EhlersR16}.
	
\item Syntactical analysis of membership in $GR(k)$ for any $k$, modulo
  boolean identities\footnote{We use the setup
    of~\cite{BloemJPPS12} to identify the transition
    structure and the $GR(k)$ winning condition.}.

\item On the fly adjustment of parameters, semantics or targets.

\item Preprocessing of the resulting LTL formula, including

\begin{itemize}

\item[$ \circ $] conversion to negation normal form,

\item[$ \circ $] replacement of derived operators, and

\item[$ \circ $] pushing/pulling next, eventually, or globally operators inwards/outwards.

\end{itemize}

\end{itemize}

\section{Extensions}
\label{sec:extensions}
The format remains open for further extensions, which allow more
fine-grained control over the specification with respect to a
particular synthesis problem. At the time of writing, the following
extensions were under consideration:

\begin{itemize}

\item Compositionality: The possibility to separate specifications
  into multiple components, which then can be used as building blocks
  to specify larger components. E.g., it should be possible to express
  the whole decomposed specification, as it is depicted in
  Figure~\ref{fig:amba} (including the interconnections), in a single
  specification file in TLSF.
	
\item Partial Implementations: a specification that is separated into
  multiple components might also contain components that are already
  implemented. Implemented components could be given in the AIGER
  format that is already used in SYNTCOMP~\cite{SYNTCOMP-format}.

\item Libraries: Several functions and definitions are often shared
  between components, e.g., the function \textit{mutual} of the example
  in \secref{sec:example}. Hence, it is more useful to ship 
  them via libraries instead of redeclaring them each time.

\item LTL Fragment Detection: Our tool currently only supports
  detection of $GR(k)$. We aim to support the detection of further
  relevant fragments, like for example Liveness or Safety.

\end{itemize}

{\small
\subsection*{Acknowledgments}
\label{sec:acknowledgments}

We thank Roderick Bloem, 
R\"udiger Ehlers, Bernd Finkbeiner, Ayrat Khalimov, Robert K\"onighofer, Nir 
Piterman, and Leander Tentrup for comments on TLSF and drafts of this 
document.

The development of TLSF has been supported by the German
Research Foundation (DFG) through project ``Automatic Synthesis of 
Distributed and
Parameterized Systems'' (JA 2357/2-1) and by the 
European Research Council (ERC) Grant OSARES (No.~683300).
}

\bibliographystyle{eptcs}
\bibliography{biblio,synthesis}

\newpage
\appendix

\section{Appendix}
\label{sec:appendix}
\subsection{Linear Temporal Logic}
\label{apx_ltl}

Linear Temporal Logic (LTL) is a temporal logic, defined over a finite
set of atomic propositions~$ \text{AP} $. The syntax of LTL conforms to the
following grammar:
\begin{equation*}
  \varphi \ \ := \ \ \text{\textit{true}} \sep  p \in \text{AP}
  \sep \neg \varphi \sep \varphi \vee \varphi
  \sep \LTLnext \varphi \sep \varphi \LTLuntil \varphi
\end{equation*}
The semantics of LTL are defined over infinite
words~$ \alpha = \alpha_{0}\alpha_{1}\alpha_{2}\dots \in
(2^{\text{AP}})^{\omega} $.
A word~$ \alpha $ satisfies a formula~$ \varphi $ at position
$ i \in \nats $:

\begin{itemize}

\item $ \alpha, i \vDash \textit{true} $ 

\item $ \alpha, i \vDash p $ \ iff \ $ p \in \alpha_{i} $

\item $ \alpha, i \vDash \neg \varphi $ \ iff \
  $ \alpha, i \not\vDash \varphi $

\item $ \alpha, i \vDash \varphi_{1} \vee \varphi_{2} $ \ iff \
  $ \alpha, i \vDash \varphi_{1} $ or $ \alpha, i \vDash \varphi_{2} $

\item $ \alpha, i \vDash \LTLnext \varphi $ \ iff \
  $ \alpha, i + 1 \vDash \varphi $

\item $ \alpha, i \vDash \varphi_{1} \LTLuntil \varphi_{2} $ \ iff \
  $ \exists n \geq i.\ \alpha, n \vDash \varphi_{2} $ and
  $ \forall i \leq j < n.\ \alpha, j \vDash \varphi_{1} $

\end{itemize}

\noindent A word~$ \alpha \in 2^{\text{AP}} $
satisfies a formula~$ \varphi $ iff $ \alpha, 0 \vDash \varphi $.
Beside the standard operators, we have the following derived operators:

\begin{itemize}

\item
  $ \varphi_{1} \wedge \varphi_{2} \equiv \neg(\neg \varphi_{1} \vee
  \neg \varphi_{2}) $

\item
  $ \varphi_{1} \rightarrow \varphi_{2} \equiv \neg \varphi_{1} \vee
  \varphi_{2} $

\item
  $ \varphi_{1} \leftrightarrow \varphi_{2} \equiv (\varphi_{1}
  \rightarrow \varphi_{2}) \wedge (\varphi_{2} \rightarrow
  \varphi_{1}) $

\item
  $ \LTLeventually \varphi \equiv \text{\textit{true}} \LTLuntil
  \varphi $

\item $ \LTLglobally \varphi \equiv \neg \LTLeventually \neg \varphi $

\item
  $ \varphi_{1} \LTLrelease \varphi_{2} \equiv \neg(\neg \varphi_{1}
  \LTLuntil \neg \varphi_{2}) $

\item
  $ \varphi_{1} \LTLweakuntil \varphi_{2} \equiv (\varphi_{1}
  \LTLuntil \varphi_{2}) \vee \LTLglobally \varphi_{1} $

\end{itemize}

\subsection{Mealy and Moore Automata}
\label{apx_memo}

A Mealy automaton is a tuple
$ \mathcal{M}_{e} = (\inputs, \outputs, Q, q_{0}, \delta, \lambda_{e})
$, where

\begin{itemize}

\item $ \inputs $ is a finite set of input letters,

\item $ \outputs $ is a finite set of output letters,

\item $ Q $ is finite set of states,

\item $ q_{0} \in Q $ is the initial state,

\item $ \delta \colon Q \times \inputs \rightarrow Q $ is the
  transition function, and

\item
  $ \lambda_{e} \colon Q \times \inputs \rightarrow \outputs $
  is the output function.

\end{itemize}

\noindent Hence, the output depends on the current state of the
automaton and the last input letter.

\bigskip

\noindent A Moore automaton is a tuple
$ \mathcal{M}_{o} = (\inputs, \outputs, Q, q_{0}, \delta, \lambda_{o})
$,
where $ \inputs, \outputs, Q, q_{0} $ and $ \delta $ are defined as
for Mealy automata. However, the output function
$ \lambda_{o} \colon Q \rightarrow \outputs $ determines the current
output only on the current state of the automaton, but not the current
input.

\subsection{TLFS encoding of Shift, TSingle, TIncr, TBurst4 and Lock}
\label{apx:components}

\subsubsection{Shift}

\vspace{0.8em}

\noindent\quad\usebox{\cShift}

\vspace{1.5em}

\subsubsection{TSingle}

\vspace{0.8em}

\noindent\quad\usebox{\cSingle}

\newpage

\subsubsection{TIncr}

\vspace{0.8em}

\noindent\quad\usebox{\cTincr}

\vspace{0.8em}

\subsubsection{TBurst4}

\vspace{0.8em}

\noindent\quad\usebox{\cBurst}

\newpage

\subsubsection{Lock}

\vspace{0.8em}

\noindent\quad\usebox{\cLock}

\end{document}